\documentclass[12pt, draftclsnofoot, onecolumn]{IEEEtran}
\IEEEoverridecommandlockouts
\usepackage[noadjust]{cite}
\usepackage{amsmath,amssymb,amsfonts}
\usepackage{graphicx}
\usepackage{textcomp}
\usepackage{setspace}
\usepackage{xcolor}
\def\BibTeX{{\rm B\kern-.05em{\sc i\kern-.025em b}\kern-.08em
		T\kern-.1667em\lower.7ex\hbox{E}\kern-.125emX}}
\usepackage{amsthm}
\usepackage[ruled,vlined]{algorithm2e}

\usepackage[noend]{algpseudocode}
\usepackage{subcaption}
\usepackage{siunitx}
\usepackage[textfont=normalfont,singlelinecheck=off,justification=raggedright,font=footnotesize]{caption}

\usepackage[margin=30pt,textfont=normalfont,singlelinecheck=off,justification=raggedright,font=footnotesize]{subcaption}
\usepackage{dsfont}

\usepackage{environ}         
\usepackage{etoolbox}        
\usepackage{graphicx}        

\usepackage{anyfontsize}         

\usepackage{xcolor}

\usepackage[hidelinks]{hyperref}
\usepackage{booktabs}
\usepackage{bbm}

\usepackage{authblk}

\usepackage{lipsum}

\newcommand\blfootnote[1]{%
	\begingroup
	\renewcommand\thefootnote{}\footnote{#1}%
	\addtocounter{footnote}{-1}%
	\endgroup
}

\newlength{\myl}
\let\origequation=\equation
\let\origendequation=\endequation

\RenewEnviron{equation}{
	\settowidth{\myl}{$\BODY$}                       
	\origequation
	\ifdimcomp{\the\linewidth}{>}{\the\myl}
	{\ensuremath{\BODY}}                             
	{\resizebox{\linewidth}{!}{\ensuremath{\BODY}}}  
	\origendequation
}

\usepackage{caption,booktabs}

\makeatletter
\def\endthebibliography{%
	\def\@noitemerr{\@latex@warning{Empty `thebibliography' environment}}%
	\endlist
}
\makeatother

\begin{document}

	\newtheorem{prop}{Proposition}
	\title{Blind Federated Learning at the Wireless Edge with Low-Resolution ADC and DAC
	}
	
	\author[1, 2]{Busra Tegin}
	\author[1]{Tolga M. Duman}

	\affil[1]{\small Department of Electrical and Electronics Engineering\protect\\Bilkent University, Ankara, Turkey}
	\affil[2]{\small Huawei Turkey Research and Development Center (HTRDC), Istanbul, Turkey}
	
	\affil[ ]{Email: {\tt btegin@ee.bilkent.edu.tr,~duman@ee.bilkent.edu.tr}}
	
	\maketitle
	\vspace{-1.5cm}
			\blfootnote{Part of the material in this paper will be submitted to
		the 2021 IEEE Global Communication Conference (GLOBECOM), Madrid,
		Spain, December 2021. }
	\begin{abstract}

		We study collaborative machine learning systems where a massive dataset is distributed across independent workers which compute their local gradient estimates based on their own datasets. Workers send their estimates through a multipath fading multiple access channel with orthogonal
		frequency division multiplexing to mitigate the frequency selectivity of the channel. We assume that there is no channel state information (CSI) at the workers, and the parameter server (PS) employs multiple antennas to align the received signals. To reduce the power consumption and the hardware costs, we employ complex-valued low-resolution digital-to-analog converters (DACs) and analog-to-digital converters (ADCs), at the transmitter and the receiver sides, respectively, and study the effects of practical low-cost DACs and ADCs on the learning performance. 
		Our theoretical analysis shows that the impairments caused by low-resolution DACs and ADCs, including those of one-bit DACs and ADCs, do not prevent the convergence of the federated learning algorithm, and the multipath channel effects vanish when a sufficient number of antennas are used at the PS.  
		We also validate our theoretical results via simulations, and demonstrate that using low-resolution, even one-bit, DACs and ADCs causes only a slight decrease in the learning accuracy.

	\end{abstract}

	\section{Introduction}
	The rapid growth of data sensing and collection capabilities of computation devices facilitates the use of massive datasets enabling machine learning (ML) systems to make more intelligent decisions than ever. 
	However, this growth makes the processing of all the data in a central processor troublesome due to increased energy  consumption and privacy concerns. 
	As an alternative to using a central processor, performing the ML task in a distributed manner, called federated learning, has recently drawn significant attention \cite{bekkerman, chilimbi}.
	In federated learning,  each device connected to the central processor performs the required gradient computation based on its local dataset, and sends it to the central processor. 
	The global parameter update is performed at the central processor using the local computations of the connected devices.

	While federated learning can be considered as a combination of two broadly studied areas: statistical learning and communications, it also opens paths for new research areas. 
	With this motivation, different problems related to federated learning are studied in the recent literature. These include studies on the effects of energy constraints, resource allocation, privacy, compression of local computations, convergence analysis of the learning algorithms, and performance over different channel models.
	In particular, in \cite{amiri1}, digital and analog distributed stochastic gradient descent (D-DSGD and A-DSGD) algorithms over a Gaussian multiple-access channel (MAC) are proposed. The authors use the superposition property of the MAC to recover the mean of the local gradients computed at remote workers. 
	In D-DSGD, workers digitally compress their locally computed gradients into a finite number of bits, while in A-DSGD, workers use an analog compression similar to what is done in compressed sensing (CS) to obey the bandwidth limitations.
	In \cite{zhu2018low} and \cite{amiri2}, the channel between the parameter server (PS) and the workers is modeled as a fading MAC. 
	Ref. \cite{zhu2018low} performs power allocation among the gradients to schedule workers according to their channel state information (CSI). 
	The authors show that the latency reduction of the proposed method scales linearly with the device population.	
	Ref. \cite{amiri2} proposes a gradient sparsification method which is followed by a CS algorithm to reduce the dimensions of large parameter vector. 
	By reducing the dimensionality of the gradients and designing a power allocation scheme, the authors obtain significant performance improvements compared to the existing benchmarks.

	In addition to the studies that decrease the communication load, Ref. \cite{yang2019energy} considers transmission energy, and formulates an optimization problem for the joint learning and communication process.
	The goal is to minimize the total energy consumption for local computations and wireless transmission under latency constraints. 
	In \cite{chen2020convergence}, the authors focus on the minimization of the convergence time of a federated learning system by jointly considering user selection and resource allocation. 
	The aim of the PS is to include as many workers as possible into the learning process for convergence to the global model with limited resources.
	%
	%
	There are also several studies on data exchange rate reduction via quantization \cite{qsgd,seide,zhao2019quantized, amiri2020federated}.
	Specifically, in \cite{amiri2020federated}, the authors introduce a lossy federated learning (LFL) system, which directly quantizes both the global and the local model parameters to reduce the communication loss. 
	They show that the convergence of the learning algorithm is guaranteed despite the quantization process.
	When the training data is randomly split among the workers, LFL with a small number of quantization levels performs as well as a system with unquantized parameters. 
	In another line of research, \cite{amiri} considers a federated learning system for which there is no CSI at the workers; hence the PS employs multiple antennas to align the received signals. 
	In \cite{amiri2020blindx}, this study is extended further, and a convergence analysis for the blind federated learning with both perfect and imperfect CSI is performed.

	While different aspects of federated learning, such as gradient compression, resource allocation, latency constraints, and fading channel effects are studied in the recent literature, the existing studies do not consider very realistic transmission models or channels. To make the use of federated learning practical, one should also consider these extensions and low-cost system design with the hardware-induced distortion for a complete system design, which is the subject of our study.

	%

	In this paper, our main objective is to study federated learning over wireless channels in realistic settings by considering practical implementation issues as well as the wireless channel effects. 
	We model the communication link as a frequency selective fading channel, and transmit the local gradients using orthogonal frequency division multiplexing (OFDM). 
	We consider the blind transmitter scenario, i.e., there is no CSI at the transmitters, hence  multiple (even a massive number of) receive antennas are employed at the receiver side. 
	Furthermore, to reduce the hardware complexity and power consumption, we employ low-resolution  digital-to-analog converters (DACs) at the transmitter side (at each worker), and analog-to-digital converters (ADCs) at the receiver side. 
	In fact, this is nothing but the over-the-air machine learning, except that here we are considering the effects of the wireless medium as well as the use of low-resolution DACs and ADCs.  
	Note that while OFDM transmission with low-resolution ADCs and DACs has extensively been studied from a communication theory perspective in the literature (see, e.g., \cite{studer2016quantized, mollen2016uplink,dardari2006joint, one0, one1, one2, xu2017user, jacobsson2017massive }), this is the first paper on their use for federated learning over wireless channels. 
	
	The main contributions of the paper can be summarized as follows: 
	\begin{itemize}
		\item Different from previous works regarding federated learning reviewed above (\cite{amiri, amiri1, amiri2, amiri2020federated, zhu2018low,qsgd,seide,zhao2019quantized, amiri2020blindx}), we consider a realistic wireless channel model where the channel between the workers and PS is modeled as a multipath fading MAC channel. 
		\item To cope with the realistic channel impairments, we transmit the local gradients using OFDM with a cyclic prefix (CP) to 
		mitigate the ISI caused by the multipath. Thus, different from \cite{amiri2020federated}, we consider the transmission and reception of actual OFDM signals, not gradients directly, as would be necessitated in a practical implementation.
		\item Since one of our main concerns is a practical implementation of federated learning, we also employ low-resolution DACs and ADCs separately at the workers and the PS side, respectively.  
		Also, we extend our studies to the case of a system which utilizes both low-resolution DACs and ADCs.
		\item Via both theoretical analysis and extensive simulations, we find that the effects of imperfections due to finite resolution DAC and/or ADC can be alleviated using a sufficient number of receive antennas at the PS, and the convergence of the distributed learning algorithm is guaranteed even if
		we employ low-cost (even one-bit) DACs and/or ADCs. 
\end{itemize}

	The paper is organized as follows. Section II introduces
	the system model and preliminaries. DSGD with low-resolution DACs is analyzed in Section III, and the  effect of low-resolution ADCs at the receiver side is studied in Section IV, respectively. 
	Joint utilization of low-resolution DACs and ADCs are considered in Section V. Performance of blind federated learning with realistic channel effects and hardware limitations is studied via simulations in Section \ref{sim}, and the paper is concluded in Section \ref{conc}.
	
	\textit{Notation:} Throughout this paper, the real and imaginary parts of $x \in \mathbb{C}$ are represented by $x^R$ and $x^I$, respectively. We use the notation $[a \ b]$ to indicate the integer set $\{a, \dots, b\}$ where $a\leq b$, $a$ and $b$ are positive integers, and $[b] = [1 \ b]$. We denote $l_2$ norm of a vector $\mathbf{x}$ by $||\mathbf{x}||_2$. The entry in the $i$-th row and $j$-th column of a matrix $\mathbf{A}$ is denoted by $\mathbf{A}[i,j]$. $N$-point Discrete Fourier Transform (DFT) of vector $\mathbf{x} \in \mathbb{C}^N$ is defined as
	\begin{equation}
	\mathbf{X}[u] = \sum_{n = 1}^{N}  \mathbf{x}[n] e^{-j2\pi n u/N}.
	\end{equation}
	while the $N$-point inverse discrete Fourier Transform (IDFT) of vector $\mathbf{X} \in \mathbb{C}^N$ is given by
	\begin{equation}
	\mathbf{x}[n] = \frac{1}{N}\sum_{u = 1}^{N}  \mathbf{X}[u] e^{j2\pi n u/N}.
	\end{equation}

	\vspace{-0.2cm}
	\section{System Model}
	\begin{figure}[t]
		\centering
		\includegraphics[trim={4cm 7cm 8cm 1cm},clip,height=0.45\textwidth]{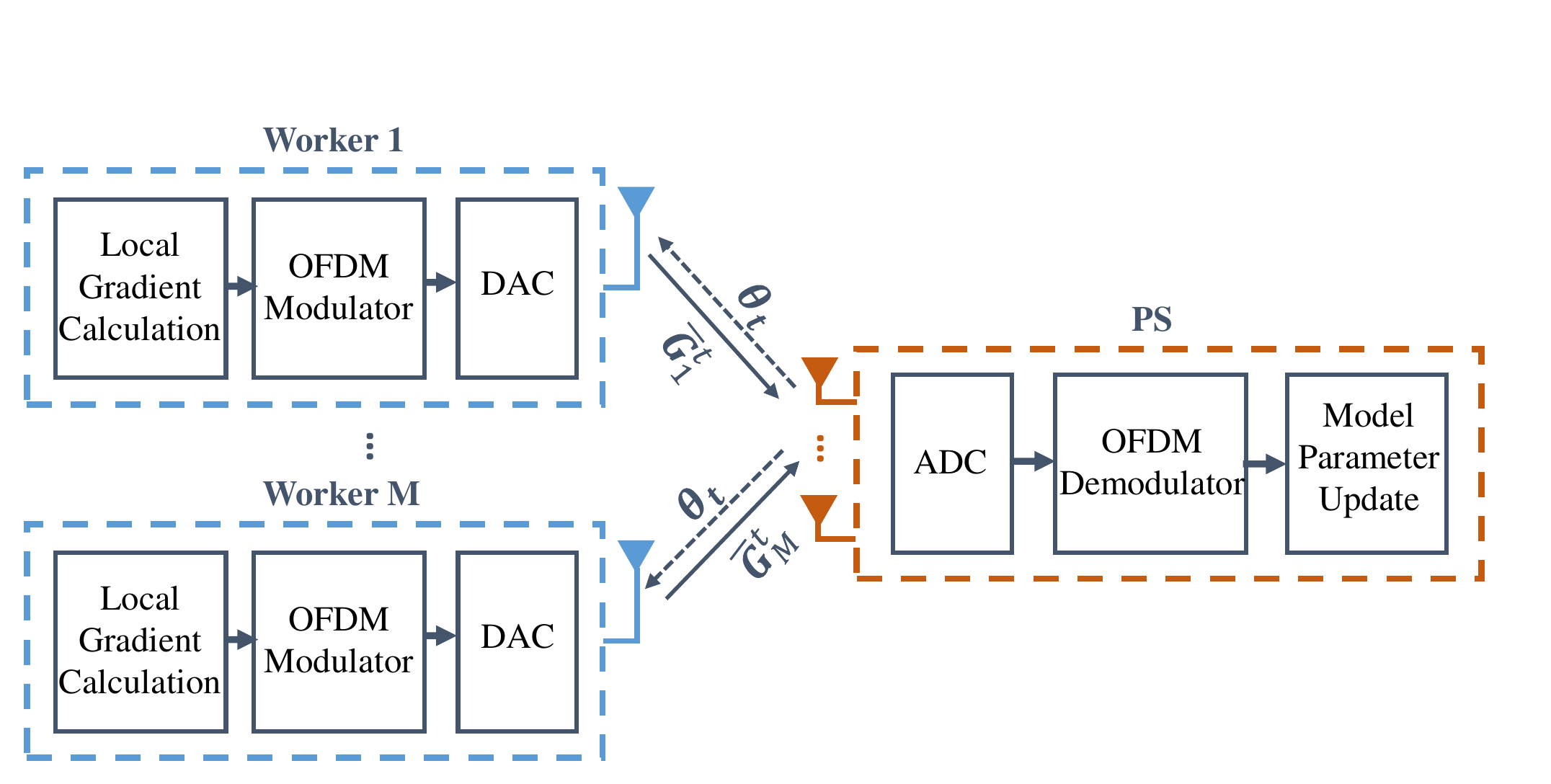}
		\caption{System model for distributed machine learning at the wireless edge.}  \label{system}
		\vspace{-0.5cm}
	\end{figure}

	We consider a distributed ML system where each worker calculates its gradient estimate and sends it to a central PS through a multipath fading MAC with OFDM as illustrated in Fig. \ref{system}. 
	At the receiver side, OFDM demodulation, signal combining and global model parameter update are performed.
	The global parameter is broadcast to the workers over an error-free link. 
	We assume that there is no transmit side CSI, and that the PS employs multiple antennas to recover the average of the workers' gradients. 
	With the use of a higher number of workers and many antennas, a significant amount of power at the transmitter and receiver is consumed by the DACs and ADCs \cite{walden1999analog}. 
	The power consumption of DACs and ADCs increases linearly, and their hardware cost increases exponentially with the number of quantization bits \cite{lee2008analog}. 
	In order to keep the implementation cost and power consumption low,  we consider a distributed learning system where the transmitters and receivers are equipped with low-resolution, even one-bit, DACs and ADCs, respectively.

	We jointly train a learning
	model by using iterative stochastic gradient descent (SGD) to minimize a loss function $f(\cdot)$. 
	During the $t$-th iteration, worker $m \in [M]$ calculates the gradient estimate  $\mathbf{g}_m^t \in \mathbb{R}^{d}$ by processing its local dataset $\mathcal{B}_m$ according to $\frac{1}{|\mathcal{B}_m|} \sum_{u \in \mathcal{B}_m} \triangledown f(\boldsymbol{\theta}_t, u)$ where $\boldsymbol{\theta}_t \in \mathbb{R}^{d}$ is the vector of model parameters, $d$ is the number of model parameters, and ${g}_m^t[n]$ represents the $n$-th entry of the gradient estimate vector.
	We form the baseband
	frequency domain signal of the local gradient vector as
	\begin{equation} \label{grad}
	\mathbf{\hat g}_m^t = \left[{ g}_m^t[1] + j { g}_m^t[s+1] , { g}_m^t[2] + j { g}_m^t[s+2], \cdots, {g}_m^t[s] + j {g}_m^t[2s]  \right],
	\end{equation}
	where $s= \lceil d/2\rceil$, $ \mathbf{\hat g}_m^t \in \mathbb{R}^{s}$, and  ${g}_m^t[2s]$ is assigned as zero if $d \equiv 1 \pmod{2}$.
	Then, the first step is to form the OFDM signal by taking an $N$-point inverse discrete Fourier Transform (IDFT) of the gradient vector as
	\begin{equation}  \label{workerG}
	{{G}}_m^t [u] = \frac{1}{N}\sum_{n = 1}^{N} \hat g_m^t[n] e^{j2\pi n u/N},
	\end{equation}
	for $u \in [N]$. If $s < N$, $\hat g_m^t[n] = 0$ for $ n > s$, i.e., $\mathbf{\hat g}_m^t$ is zero padded.

	The channel between the $m$-th worker and the $k$-th antenna of the PS is modeled as a (wireless) multipath MAC. 
	We assume that the channel does not change during the transmission of one OFDM word, while it may be different for different OFDM words.  
	The impulse response of the channel is 
	\begin{equation} \label{ch_impulse}
	h_{mk}^t[n] = \sum_{l = 1}^L h_{mkl}^t \delta[n-\tau_{mkl}],
	\end{equation}
	where $n \in [N+N_{cp}]$, $L$ is the number of channel taps, $\tau_{mkl}$ is the time delay and $h_{mkl}^t \in \mathbb{C}$ is the gain of the $l$-th channel tap from the $m$-th worker to the $k$-th antenna of the PS. 
	Note that this is nothing but the machine learning over-the-air framework. 
	We assume that $h_{mkl}^t$ are zero-mean complex Gaussian with $\mathbb{E}\left[(h_{mkl}^t) \cdot (h_{m'k'l'}^t)^*\right] = 0$ for $(m,k,l) \neq (m',k',l')$, and $\mathbb{E}\left[|h_{mkl}^t|^2\right] = \sigma^2_{h,l}$, i.e., all the channel taps experience Rayleigh fading.

	To mitigate the ISI caused by the multipath channel, CP addition is performed by  
	\begin{equation} \label{ofdm_words}
	\mathbf{\bar{G}}_m^t = \big[{G}_m^t[N-N_{cp}+1] \dots  {G}_m^t[N] \ {G}_m^t[1]  \dots {G}_m^t[N]\big],
	\end{equation}
	where $\mathbf{\bar{G}}_m^t \in \mathbb{C}^{N+N_{cp}}$ is the OFDM word to be transmitted by the $m$-th worker. 
	The CP length $N_{cp}$ is chosen to be greater than the delay spread of all the channels.	
	The resulting (depending on the setup -- quantized or full resolution) OFDM words are transmitted to the PS which are equipped with $K$ receive antennas. 
	The PS uses the received signal to update the model and sends it back to all the receivers  over an error-free link.

	At the $k$-th receive chain, after removing the CP, the $n$-th entry of the received vector at the input of the $k$-th receive antenna during iteration $t$ is written as 
	\begin{equation} \label{Yk}
	Y_k^t[n] = \sum_{m=1}^M \sum_{l=1}^L h_{mkl}^t G_m^t[n-\tau_{mkl}] + z_k^t[n].
	\end{equation}
	We model the communication link as a frequency selective fading channel whose impulse response is given in \eqref{ch_impulse}, and $G_m^t[n]$ is the transmitted signal by the $m$-th worker. 
	The additive noise terms $z_k^t[n]$ are independent and identically distributed (i.i.d.) circularly symmetric zero mean complex Gaussian random variables, i.e., $z_k^t[n] \sim \mathcal{CN}(0,\,\sigma_z^2)$ for $k \in [K]$.

	Ideally, the PS updates the model parameter according to $\boldsymbol{\theta}_{t+1} = \boldsymbol{\theta}_{t} - \mu_t \frac{1}{M} \sum_{m=1}^M  \mathbf{g}_m^t$, and it is shared with the workers. 
	However, in our setup, the local gradients are not available at the PS, instead the PS uses noisy and corrupted version (by low-resolution DAC and/or ADCs) of the local gradients to recover the estimate of the gradient vector as will become apparent in the subsequent sections. 
	In the following, we drop the
	subscripts referring to iteration count $t$ for ease of exposition.

	\section{DSGD with Low-Resolution DACs at the Workers} \label{fin_dacs}
	In this section, we study the effects of employing low-resolution DACs at the workers on the distributed learning process in an effort to reduce the hardware complexity and power consumption. 
	
	After constructing the OFDM word corresponding to the gradient vectors, a complex-valued low-resolution DAC is employed to generate the transmitted signal at each worker. 
	A $b$-bit complex-valued DAC consists of two parallel real-valued DACs with quantization function ${Q_b}(\cdot)$.
	The real and imaginary parts are separately quantized into $\beta = 2^b$ reconstruction levels. 
	The reconstruction levels are denoted by $\mathbf{\hat{a}} = [\hat{a}_1 \ \hat{a}_2 \cdots \hat{a}_{\beta}] \in \mathbb{R}^\beta$ while the boundaries of the quantization regions are denoted by $\mathbf{\hat{x}} = [\hat{x}_1 \ \hat{x}_2 \cdots \hat{x}_{\beta+1}] \in \mathbb{R}^{\beta+1}$ where $\hat{x}_1 = -\infty$ and $\hat{x}_{\beta+1} = +\infty$ for convenience. 
	Also, we have, $\hat{a}_i < \hat{a}_j$, if $1 \leq i < j \leq \beta$, $\hat{x}_i < \hat{x}_j$ if $1 \leq i < j \leq \beta+1$, and $\hat{x}_i \leq \hat{a}_j < \hat{x}_k$ if $1 \leq i \leq j < k \leq \beta+1$. 
	The corresponding real valued quantizer is $Q_b(z) = \hat{a}_i$ for $\hat{x}_i \leq z < \hat{x}_{i+1}$,
	$i \in [\beta]$, $z \in \mathbb{R}$. The complex-valued DAC operation can be expressed as
	$
	Q_b( x ) = Q_b(x^R) + jQ_b(x^I)
	$. 
	We assume that the quantizer output
	is chosen such that $Q_b(x) = \mathbb{E}[{X} |Q_b({X})]$, i.e., the reconstruction level is selected to minimize the mean squared error for each quantization region. 
	The corresponding signal to quantization noise ratio (SQNR) of the input vector $\mathbf{x}$ is calculated as
	\begin{equation}
	\text{SQNR} = \frac{\mathbb{E}\left[|{{X}}|^2\right]}{\mathbb{E}\left[|Q_b({{X}})-{{X}}|^2\right]} = 1-\frac{\mathbb{E}\left[Q_b({X}){X}^*\right]}{\mathbb{E}\left[|Q_b({X})-{X}|^2\right]}.
	\end{equation}

	    \begin{figure}[t]
		\centering
		\includegraphics[trim={1.2cm 0 0 0},clip,width=0.7\textwidth]{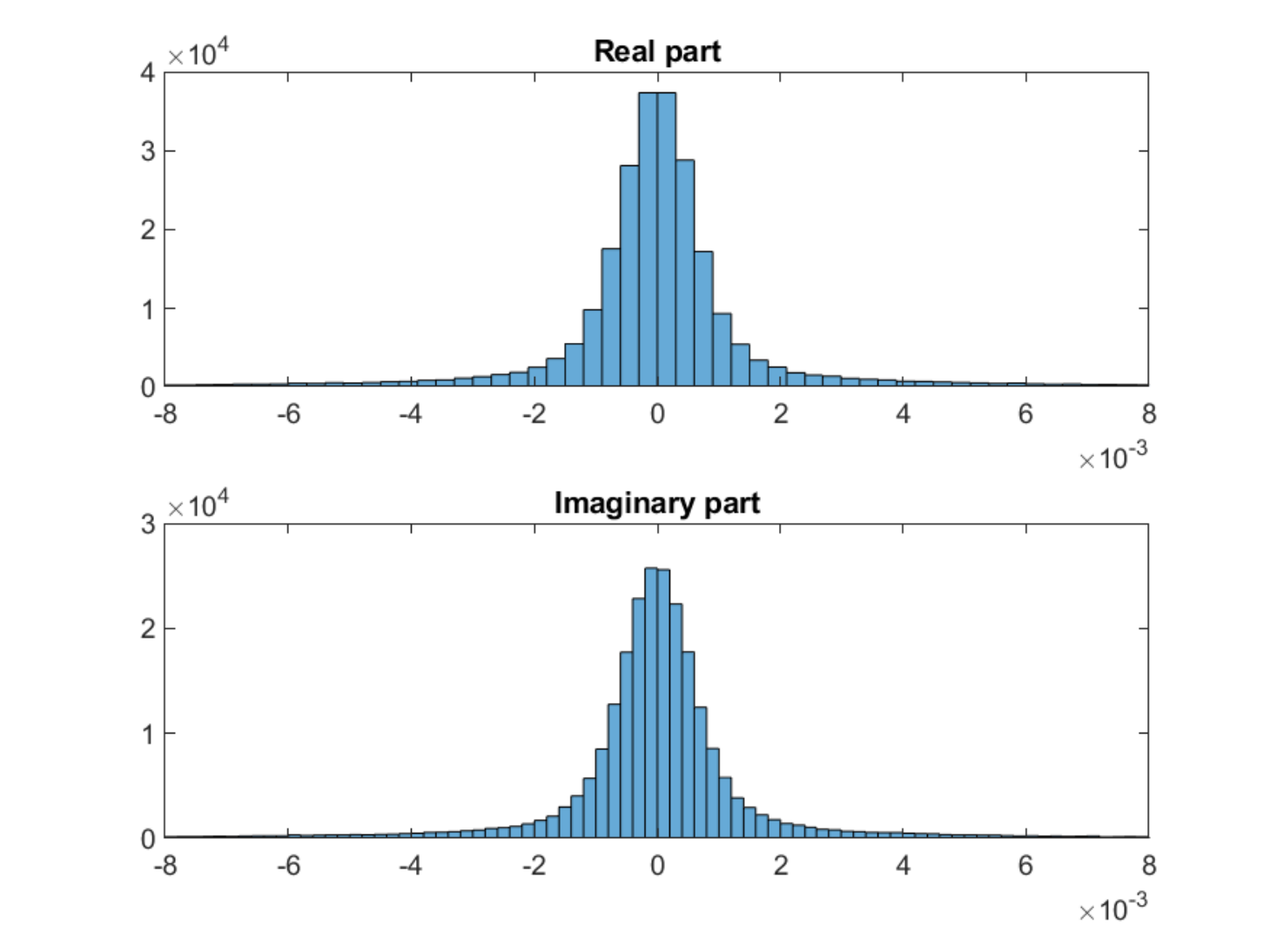}
		\caption{Histogram of the real and imaginary parts of the OFDM word.}  \label{hist2}
	\end{figure}
	
	We model the OFDM words as wide-sense stationary (WSS) Gaussian processes based on an argument similar to the one made in \cite{ofdm}. That is, if the input data which forms the OFDM word is i.i.d. and bounded, the convex envelope of the OFDM word weakly converges to a Gaussian random process as the number of subcarriers goes to infinity through an application of central limit theorem (CLT). 
	Similarly, if we assume that the elements of the gradient vector in the learning process are i.i.d. and bounded, then the real and imaginary parts of the baseband OFDM word obtained from the gradient vector can be modeled as independent zero-mean stationary Gaussian processes. 
	As a verification, we examine histograms of several OFDM word samples obtained by a certain learning task with our setup, demonstrating the OFDM samples are approximately Gaussian. 
	An instance of an exemplary histogram of the OFDM word samples obtained through the $100$-th iteration is given in Fig. \ref{hist2} which is consistent with our assumption. 
	Our extensive experiments further confirm that the corresponding OFDM word samples at different time indexes have almost the same variance. 
	Note that, even if the OFDM words are not Gaussian processes, the Bussgang theorem that will be used to model the nonlinear input-output relationship for DACs and ADCs is still a good approximation as illustrated extensively in the literature, see, e.g.,
	\cite{jacobsson2018massive}-\cite{aghdam2019performance}.

		\begin{table}[t]
		\centering
		\caption{Distortion factors with different quantization levels \cite{fan2015uplink,max1960quantizing}.}  \label{tab1}
		\begin{tabular}{|c|c|}
			\hline
			\textbf{Number of bits} & \textbf{Distortion factor ($\eta$)} \\ \hline
			1                       & 0.3634                              \\ \hline
			2                       & 0.1175                              \\ \hline
			3                       & 0.03454                             \\ \hline
			4                       & 0.009497                            \\ \hline
			5                       & 0.002499                            \\ \hline
		\end{tabular}
	\end{table}

	We denote the autocorrelation matrix of the OFDM words by $\mathbf{C}_{\mathbf{\bar{G}}_m \mathbf{\bar{G}}_m}$ with  equal diagonal elements denoted by $\sigma_{G_m}^2$. 
	Using the Bussgang decomposition \cite{bussgang}-\cite{demir2020bussgang}, we can decompose the quantized signal into two parts: the desired signal component and the quantization distortion which is uncorrelated of the desired signal. 
	Thus, we can write the quantized signal as
	\begin{equation} \label{aqnm_eq}
	{\bar{G}}_m^Q[n] = Q({\bar{G}}_m[n]) = (1-\eta){\bar{G}}_m[n] + {q_m[n]},
	\end{equation}
	where $\eta = 1/\text{SQNR}$ is the distortion factor which is the inverse of SQNR, and the variance of the distortion noise is $\sigma_{q_m}^2 = \eta (1-\eta) \sigma_{G_m}^2$. 
	When a unit variance Gaussian input is processed by a non-uniform scalar
	minimum mean-square-error quantizer, the values of corresponding distortion factors are listed in Table \ref{tab1} \cite{fan2015uplink}-\cite{max1960quantizing}.

	 At the $k$-th receive chain, after removing the CP, the $n$-th entry of the received vector is written as 
	 \begingroup
	 \allowdisplaybreaks
	\begin{align} \label{Yk1}
	Y_k[n] &= \sum_{m=1}^M \sum_{l=1}^L h_{mkl} G_m^Q[n-\tau_{mkl}] + z_k[n] \\
	&= \sum_{m=1}^M \sum_{l=1}^L h_{mkl} \left( (1-\eta)\cdot G_m[n-\tau_{mkl}] + q_m[n-\tau_{mkl}] \right)+ z_k[n] \\
	&= (1-\eta) \sum_{m=1}^M \sum_{l=1}^L  h_{mkl}  G_m[n-\tau_{mkl}] + w_{k}[n],
	\end{align}
	\endgroup
	where the total non-Gaussian noise term $w_k[n]$ has variance  $\sigma_z^2 +\eta(1-\eta)\sigma_{G_m}^2\sum_{m=1}^M \sum_{l=1}^L |h_{mkl}|^2$.

	To perform the demodulation, we take the DFT of (\ref{Yk1}) which gives
	\begin{equation} \label{fftyqk1}
	r_k[i] = (1-\eta) \sum_{m=1}^M H_{mk}[i] g_m[i] +  \sum_{m=1}^M H_{mk}[i] Q_m[i] + Z_k[i],
	\end{equation}
	where  $Q_m[i]$ is the DFT of the quantization distortion noise and 
	$H_{mk}[i]$'s are the channel gains from the $m$-th worker to the $k$-th receive chain for the $i$-th subcarrier. 
	$H_{mk}[i]$'s are given by
	\begin{align} \label{Hmk}
	H_{mk}[i] &= \sum_{n = 0}^{N-1} h_{mk}[n]e^{-j2\pi i n/N} \nonumber \\
	&= \sum_{n = 0}^{N-1} \left(\sum_{l = 1}^L h_{mkl} \delta[n-\tau_{mkl}]\right)e^{-j2\pi i n/N}  \nonumber \\
	&= \sum_{l = 1}^L h_{mkl}e^{-j2\pi i \tau_{mkl}/N}.
	\end{align}
	
	Since the channel taps are zero mean circularly symmetric complex Gaussian (i.e., Rayleigh fading), $H_{mk}[i]$'s are also zero-mean complex Gaussian random variables with variance $\sigma_{H}^2 = \sum_{l=1}^L \sigma_{h,l}^2$.
	
	Taking DFT of the channel noise vector, $Z_k[i]$ is evaluated as
	\begin{equation} \label{Zk}
	Z_k[i] = \sum_{n = 0}^{N-1} z_k[n]e^{-j2\pi i n/N}. 
	\end{equation}
	The noise terms are i.i.d. circularly symmetric complex
	Gaussian, i.e., $Z_k[n] \sim \mathcal{CN}(0,\,\sigma_{Z_k}^2)$ where $\sigma_{Z_k}^2 = N \sigma_{z_k}^2$.

	We assume that the CSI is available at the PS, hence the received signals from the $K$ antennas can be combined to align the gradient vectors using
	\begin{equation} \label{yQ1}
	y[i] = \frac{1}{(1-\eta) \cdot K}\sum_{k=1}^K \bigg(\sum_{m=1}^M (H_{mk}[i])^*\bigg) r_k [i],
	\end{equation}
	as in \cite{amiri}.	
	By substituting (\ref{fftyqk1}) into (\ref{yQ1}), we obtain
	\begingroup
	\allowdisplaybreaks
	\begin{subequations} \label{y_all11}
		\begin{align} 
		y [i] =& \underbrace{\frac{1}{K} \sum_{k=1}^K \sum_{m=1}^M |H_{mk}[i]|^2g_m[i]}_{\text{signal term}} \\  
		&+ \underbrace{\frac{1}{K} \sum_{k=1}^K \sum_{m=1}^M \sum_{\substack{m'=1 \\ m' \neq m}}^M (H_{mk}[i])^* H_{m'k}[i] g_{m'}[i]}_{\text{interference term}} \\  
		&+ \underbrace{\frac{1}{(1-\eta) K} \sum_{k=1}^K \sum_{m=1}^M \sum_{\substack{m'=1 \\ m' \neq m}}^M (H_{mk}[i])^* H_{m'k}[i] Q_{m'}[i]}_{\text{distortion noise term}} \label{y_allc} \\  
		&+ \underbrace{\frac{1}{(1-\eta) K} \sum_{k=1}^K \sum_{m=1}^M  |H_{mk}[i]|^2 Q_{m}[i]}_{\text{second type of distortion noise term }}  \label{y_alld}  \\  
		&+ \underbrace{ \frac{1}{(1-\eta) K} \sum_{k=1}^K  \bigg(\sum_{m=1}^M (H_{mk}[i])^*\bigg) Z_k[i]}_{\text{channel noise term}}.  
		\end{align}
	\end{subequations}
	\endgroup
	
	There are five different terms in (\ref{y_all11}): the signal component, interference, distortion noise term, the second type of distortion noise term, and the channel noise. 
	
	To analyze the interference term (\ref{y_all11}b), we write it as a summation of $M$ terms
		\begin{align}
\frac{1}{K} \Bigg[ &\Big(\sum_{k=1}^K \sum_{m=2}^M  (H_{mk} [i])^* H_{1k}[i] \bigg)g_{1}[i] + \cdots  \nonumber \\ & \ \ \ \ \ \ \ + \bigg(\sum_{k=1}^K \sum_{\substack{m=1 \\ m \neq j}}^M  (H_{mk} [i])^* H_{jk}[i] \bigg)g_{j}[i] + \cdots \nonumber \\
&\ \ \ \ \ \ \  + \bigg(\sum_{k=1}^K \sum_{m=1}^{M-1}  (H_{mk} [i])^* H_{Mk}[i] \bigg)g_{M}[i] \Bigg],
\end{align}
	and consider the coefficient of each term $g_{j}[i]$ separately. 
	Let us define
	\begin{equation} \label{kappa}
	\kappa_j[i] = \frac{1}{K} \sum_{k=1}^K \sum_{\substack{m=1 \\ m \neq j}}^M (H_{mk}[i])^* H_{jk}[i],
	\end{equation}
	for the coefficient of the $j$-th interfering gradient $g_{j}[i]$ in (\ref{y_all11}b) where $i \in [N]$, and $j \in [M]$. Since $H_{mk}[i]$ and $H_{jk}[i]$ are independent for $j \neq m$, the mean and variance of $\kappa_j[i]$ are calculated as
	\begin{subequations} \label{kapa}
		\begin{align}
		\mathbb{E}\left[\kappa_j[i]\right] &= 0, \\
		\mathbb{E}\left[|\kappa_j[i]|^2\right] &= \frac{(M-1)\sigma_H^4}{K}.
		\end{align}
	\end{subequations}
	We have $M$ such interference terms in (\ref{y_all11}b) each for a different worker with zero mean, and variance scaling with $\frac{M-1}{K}$. Hence, all of $M$ interference terms approach zero as $K \rightarrow \infty$. 
	
	To analyze the distortion noise term  (\ref{y_all11}c), we define the coefficient of each uncorrelated distortion term $Q_{j}[i]$ separately as in the case of (\ref{y_all11}b) by
	\begin{equation} \label{delta1}
	\delta_{1j}[i] = \frac{1}{(1-\eta) K} \sum_{k=1}^K \sum_{\substack{m=1 \\ m \neq j}}^M   (H_{mk}[i])^* H_{jk}[i],
	\end{equation}
	where $i \in [N]$, and $j \in [M]$ for all uncorrelated $M$ terms in the summation (\ref{y_all11}c).
	
	Similar to the analysis of $\kappa_j[i]$, the mean and variance of $\delta_{1j}[i]$ are calculated as
	\begin{subequations} \label{delta1}
		\begin{align}
		\mathbb{E}\left[\delta_{1j}[i]\right] &= 0, \\
		\mathbb{E}\left[|\delta_{1j}[i]|^2\right] &= \frac{{(M-1)}\sigma_H^4}{(1-\eta)^2K}.
		\end{align}
	\end{subequations}
	This implies that each of the $M$ interfering terms in (\ref{y_all11}c) goes to zero if $K$ is large enough. Thus, the detrimental effect of the distortion noise term can also be eliminated by employing a large number of receive antennas.

	To analyze the second type of distortion noise term (\ref{y_all11}d), we consider each distortion interference $Q_{j}[i]$ separately for $j \in [M]$, and define the coefficient of the interfering distortion term caused by the $j$-th one as
	\begin{equation} \label{delta2}
	\delta_{2j}[i] = \frac{1}{(1-\eta)K} \sum_{k=1}^K |H_{jk}[i]|^2 ,
	\end{equation}
	where $i \in [N]$, and $j \in [M]$.
	The mean of $\delta_{2j}[i]$ is 
	\begin{equation}
	\mathbb{E}\left[\delta_{2j}[i]\right] = \frac{ \sigma_H^2 }{(1-\eta)}.
	\end{equation}
	
	For the variance of $\delta_{2j}[i]$, we have 
	\begin{equation}
	\mathbb{E}\left[\left|\delta_{2j}[i]\right|^2\right]  = \frac{1}{(1-\eta)^2K^2} \sum_{k_1=1}^K \sum_{k_2=1}^K  \mathbb{E}\left[\left|H_{jk_1}[i]\right|^2 |H_{jk_2}[i]|^2\right].
	\end{equation}
	\begin{itemize}
		\item If  $k_1 = k_2$ (case 2.1)
		\begin{equation}
		\mathbb{E}\left[|\delta_{2j}[i]|^2\right] \big|_{\text{case 2.1}} = \frac{1}{(1-\eta)^2K^2} \sum_{k=1}^K   \mathbb{E}\left[|H_{jk}[i]|^4 \right]= \frac{1}{(1-\eta)^2K} \mathbb{E}\left[|H_{jk}[i]|^4 \right].
		\end{equation}
		
		\item If  $k_1 \neq k_2$ (case 2.1)
		\begin{align}
		\mathbb{E}\left[|\delta_{2j}[i]|^2\right] \big|_{\text{case 2.2}} 
		&= \frac{1}{(1-\eta)^2K^2} \sum_{k_1=1}^K \sum_{\substack{k_2=1 \\ k_2 \neq k_1}}^K
		 \mathbb{E}\left[ |H_{jk_1}[i]|^2 \right] \mathbb{E}\left[|H_{jk_2}[i]|^2 \right] \nonumber \\
		&= \frac{(K^2-K) \sigma_{H}^4}{(1-\eta)^2K^2}\\
		&\approx \frac{\sigma_{H}^4}{(1-\eta)^2},
		\end{align}
	\end{itemize}
	for $K \gg 1$. Thus, the mean and variance of the second distortion term of the $j$-th worker is calculated as
\begin{subequations} \label{delta2_mv}
	\begin{align}
	\mathbb{E}\left[\delta_{2j}[i]\right] &= \frac{ \sigma_H^2 }{(1-\eta)}, \\
	\text{Var}(\delta_{2j}[i]) &\approx \frac{1}{(1-\eta)^2K} \mathbb{E}\left[|H_{jk}[i]|^4 \right].
	\end{align}
\end{subequations}
Note that $\delta_{2j}[i]$ has a finite mean and its variance approaches zero as $K \rightarrow \infty$.
We know that the mean of the distortion term, $Q_{j}[i]$ for all $j \in [M]$, is zero. Accordingly, using the law of large numbers, the summation will converge to the mean of $Q_{j}[i]$, which is zero, for a sufficiently large $M$.

Using the law of large numbers, as the number of antennas at the PS $K \rightarrow \infty$, the signal term can be approximated as
\begin{equation}
y_{\text{sig}} [i] = \sigma_H^2 \sum_{m=1}^M g_m[i].
\end{equation}
Thus, with low-resolution DACs at the workers, the PS can recover the $i$-th entry of the desired signal using
\begin{align} \label{recover}
\frac{1}{M} \sum_{m=1}^M  g_m[i] = 
\begin{cases}
\frac{y^R [i]}{M\sigma_H^2},& \text{if } 1 \leq i \leq s,\\
\frac{y^I [i-s]}{M\sigma_H^2},              & \text{if } s < \ i \leq 2s.
\end{cases}
\end{align}

This result clearly shows that the destructive effect of low-resolution DACs can be effectively alleviated using a sufficient number of PS antennas. 
Thus, the convergence of the learning process is guaranteed even if
we employ low-cost low-resolution DACs at the workers, which significantly reduces the cost of designing distributed learning systems with a high number of workers. 
On the other hand, using a very large number of PS antennas will increase both the design cost and energy consumption, hence it may not be efficient. 
For further assessment, we can consider the coefficients of the distortion terms. 
For the distortion noise term given in \eqref{y_allc}, we have $M$ contributing terms each with zero mean and variance $\frac{{(M-1)}\sigma_H^4}{(1-\eta)^2K}$. To reduce the effects of these terms on the learning accuracy, it is desired to have this variance close to zero.
Clearly, this variance depends on several parameters, hence, to evaluate the overall performance, we should not only consider the number of receive antennas $K$, but also the channel variance $\sigma_H^2$, number of workers $M$, and distortion factor $\eta \in [0, \ 1]$. 
For example, if we have a high-resolution DAC, $\eta$ will be small; hence, using a smaller number of receive antennas may be sufficient to cancel out the resulting impairments. 
However, when the resolution is very low, e.g., for a one-bit DAC, $\eta$ will be large, and we will need a higher number of receive antennas due to $\frac{1}{(1-\eta)^2}$ term. 
A similar approach can also be used to analyze the second type of distortion noise term given in \eqref{y_alld} for which we have $M$ contributing terms each with variance $\frac{1}{(1-\eta)^2K} \mathbb{E}\left[|H_{jk}[i]|^4 \right]$. In other words, there is a trade-off between the DAC resolution and the number of receive antennas, and the overall performance is also affected by the channel statistics.

\section{DSGD with Low-Resolution ADCs at the PS} \label{fin_adcs}
In this section, we consider a system where the workers transmit the OFDM words corresponding to the local gradients with full-resolution through a multipath fading channel while the PS employs low-resolution ADCs at each receive antenna, and analyze  the convergence of the federated learning algorithm.

At each receive chain, after removing the CP, the $n$-th entry of the received OFDM word $\mathbf{Y_k}$ is
\begin{equation}\label{Yk_fi_res}
Y_k[n] = \sum_{m=1}^M \sum_{l=1}^L h_{mkl} G_m[n-\tau_{mkl}] + z_k[n].
\end{equation}

The ($k$, $k'$)-th element of the auto-correlation matrix of $\mathbf{{Y}}[n] = \left[Y_1[n] \cdots Y_K[n]\right] $ received by different antennas can be written as
\begin{align} \label{varY_new}
\mathbf{{C}_{YY}}[k,k'] &= \mathbb{E} \left[ \sum_{m=1}^M  \sum_{m'=1}^M \sum_{l =1}^L \sum_{l' =1}^L h_{mkl} h_{m'k'l'}^* G_m[n-\tau_{mkl}] G_{m'}^*[n-\tau_{m'k'l'}] \right] + \sigma_z^2  \mathbbm{1}_{\{k = k'\}} \\
&= \sum_{m=1}^M  \sum_{l =1}^L \sum_{l' =1}^L  h_{mkl} h_{mk'l'}^* \mathbb{E} \left[G_m[n-\tau_{mkl}] G_{m'}[n-\tau_{mk'l'}] \right] + \sigma_z^2  \mathbbm{1}_{\{k = k'\}},
\end{align}
which is only a function of $k$ and $k'$ since the OFDM words are modeled as WSS.

The variance of the received signal at the $k$-th antenna 
 $Y_k[n]$ is given by
\begin{align} \label{varY}
\sigma_{Y_k}^2 &= \mathbb{E} \left[ \sum_{m=1}^M  \sum_{m'=1}^M \sum_{l =1}^L \sum_{l' =1}^L h_{mkl} h_{m'k'l'}^* G_m[n-\tau_{mkl}] G_{m'}^*[n-\tau_{m'k'l'}] \right] +\sigma_z^2 \\
&= \sum_{m=1}^M  \sum_{l =1}^L \sum_{l' =1}^L  h_{mkl} h_{mkl'}^* \mathbb{E} \left[ G_m[n-\tau_{mkl}] G_{m}^*[n-\tau_{mkl'}] \right] +\sigma_z^2,
\end{align}
which only depends on $k$. 

A complex-valued low-resolution ADC employed at each receive antenna performs quantization.
As in the case with low-resolution DACs described in the previous section, we describe $b$-bit  quantization with quantization function ${Q_b}(\cdot)$ that independently quantizes the
real and imaginary parts into $\beta = 2^b$ reconstruction levels such that the quantizer output is chosen as $Q_b(x) = \mathbb{E}[{X} |Q_b({X})]$.

With element-wise quantization, we can decompose the quantized signal into two parts as the desired signal component and quantization distortion which is uncorrelated with the desired signal. 
Analytically, we can write the quantized signal as
\begin{equation} \label{Yk2}
R_k[n]  = (1-\eta_k) \bigg(\sum_{m=1}^M \sum_{l=1}^L h_{mkl} G_m[n-\tau_{mkl}] + z_k[n]\bigg) 
+ w_q^k[n],
\end{equation}
where $\eta_k$ is the distortion factor which is the inverse of the SQNR due to quantization of $\mathbf{{Y}}_k$. 
To determine $\eta$, one can use Table \ref{tab1}. 
$w_q^k[n]$ is a non-Gaussian distortion noise at the $k$-th antenna whose variance is  $\sigma_{w_q^k}^2 = \eta_k (1-\eta_k) \sigma_{Y_k}^2$.

The receive antennas at the PS are equipped with identical ADCs. 
As explained in \cite{demir2020bussgang}, while it may be tempting to think that the quantization noise at different ADCs is uncorrelated, this is generally not the case since each antenna receives different (delayed) linear combinations of the same set of OFDM words generated at the workers. 
On the other hand, as shown in \cite{bjornson2018hardware}, the distortion can be safely approximated as uncorrelated for massive MIMO systems with a sufficient number of users. 
We have also validated this approximation for our system, and observed that the correlation across the antennas of the PS is near-zero, even for the one-bit ADC case. 
Therefore, the correlations can be ignored as in the additive quantization noise model (AQNM), leading to a tractable scheme \cite{aqnm}. 
We further note that there are different studies on low-resolution ADCs which also neglect the distortion correlation among antennas as in our approach \cite{fan2015uplink, orhan2015low}-\cite{zhang2018mixed}. 
For zero-mean Gaussian processes, this approach is equivalent to the Bussgang decomposition, except that it ignores the correlation among the elements of the distortion term.



If we define the total effective noise due to the channel and the quantization process as
\begin{equation}
w_k[n] = (1-\eta_k) z_k[n] + w_q^k[n],
\end{equation}
the output of the complex ADC can be written as
\begin{equation} \label{Yqk_fr}
R_k[n] = (1-\eta_k) \sum_{m=1}^M \sum_{l=1}^L h_{mkl} G_m[n-\tau_{mkl}]  + w_k[n],
\end{equation}
where $w_k[n]$ is non-Gaussian total noise with variance $\sigma_{w_k}^2 = \sigma_{w_q^k}^2 + (1-\eta_k)^2\sigma_z^2 $, and assumed to be uncorrelated across the antennas.
%

To perform the OFDM demodulation, we take the DFT of (\ref{Yqk_fr}) which results in
\begin{equation} \label{fftyqk_fr}
r_k[i] = (1-\eta_k) \sum_{m=1}^M H_{mk}[i] g_m[i]  + W_k[i],
\end{equation}
where  $H_{mk}[i]$'s are the channel gains from the $m$-th worker to the $k$-th receive chain for the $i$-th subcarrier, given by \eqref{Hmk} which are zero-mean Gaussian random variables with variance $\sigma_{H}^2 = \sum_{l=1}^L \sigma_{h,l}^2$.

Taking the DFT of the effective noise, $W_k[i]$ is evaluated as
\begin{equation} \label{Wk}
W_k[i] = \sum_{n = 0}^{N-1} w_k[n]e^{-j2\pi i n/N}. 
\end{equation}
We know that the channel noise is i.i.d., and we assume that the distortion noise is $m$-dependent to decorrelate fast enough, i.e., $m \ll N$. Hence, $W_k[i]$ converges absolutely to a Gaussian random variable by an application of the central limit theorem (CLT) \cite{billingsley}, i.e., $W_k[n] \sim \mathcal{CN}(0,\,\sigma_{W_k}^2)$ where $\sigma_{W_k}^2 = N \sigma_{w_k}^2$.

Assuming that the CSI is available at the PS as in the previous section, the received signals from the $K$ antennas can be combined to align the gradient vectors by
\begin{equation} \label{yQ_fr}
y[i] = \frac{1}{K}\sum_{k=1}^K \frac{1}{1-\eta_k}\bigg(\sum_{m=1}^M (H_{mk}[i])^*\bigg) r_k [i].
\end{equation}
By substituting (\ref{fftyqk_fr}) into (\ref{yQ_fr}), we obtain
\begin{subequations} \label{y_all_fr}
	\begin{align} 
	y [i] =& \underbrace{\frac{1}{K} \sum_{k=1}^K \sum_{m=1}^M |H_{mk}[i]|^2g_m[i]}_{\text{signal term}} \\  
	&+ \underbrace{\frac{1}{K} \sum_{k=1}^K \sum_{m=1}^M \sum_{m'=1, m' \neq m}^M (H_{mk}[i])^* H_{m'k}[i] g_{m'}[i]}_{\text{interference term}} \\  
	&+ \underbrace{ \frac{1}{K} \sum_{k=1}^K \frac{1}{1-\eta_k} \bigg(\sum_{m=1}^M (H_{mk}[i])^*\bigg) W_k[i]}_{\text{noise term}}.  
	\end{align}
\end{subequations}

There are three different terms in (\ref{y_all_fr}): the signal component, interference and noise. Using the law of large numbers, as the number of antennas at the PS $K \rightarrow \infty$, the signal term approaches
\begin{equation}
y_{\text{sig}} [i] = \sigma_H^2 \sum_{m=1}^M g_m[i].
\end{equation}
Thus, the PS can recover the $i$-th entry of the desired signal
\begin{equation}
\frac{1}{M} \sum_{m=1}^M g_m[i] = \frac{y_{\text{sig}} [i]}{M\sigma_H^2}.
\end{equation} 
To analyze the interference term (\ref{y_all_fr}b), we follow the same approach as in the previous section where each $M$ interfering terms are analyzed separately. 
We define the term due to the $j$-th interfering worker as
\begin{equation} \label{kappa_fr}
\kappa_j[i] = \frac{1}{K} \sum_{k=1}^K \sum_{\substack{m=1 \\ m \neq j}}^M (H_{mk}[i])^* H_{jk}[i],
\end{equation}
where $i \in [N]$, and $j \in [M]$. Since $H_{mk}[i]$ and $H_{jk}[i]$ are independent for $j \neq m$, the mean and variance of $\kappa_j[i]$ are calculated as
\begin{subequations} \label{kapa_fr}
	\begin{align}
	\mathbb{E}\left[\kappa_j[i]\right] &= 0, \\
	\mathbb{E}\left[|\kappa_j[i]|^2\right] &= \frac{(M-1)\sigma_H^4}{K}.
	\end{align}
\end{subequations}
Accordingly, for fixed gradient values, each of the $M$ interference terms in (\ref{y_all_fr}b) has zero mean and their variances scale with $\frac{M-1}{K}$. Thus, similar to the ideal case (where the receive chains are equipped with infinite resolution ADCs as considered in \cite{amiri}), the interference term approaches zero as $K \rightarrow \infty$. In other words, using a sufficiently large number of antennas at the PS eliminates the destructive effects of the interference on the learning process, and the estimate for the gradient vector is obtained as
\begin{align}
\frac{1}{M} \sum_{m=1}^M  g_m[i] = 
\begin{cases}
\frac{y^R [i]}{M\sigma_H^2},& \text{if } 1 \leq i \leq s,\\
\frac{y^I [i-s]}{M\sigma_H^2},              & \text{if } s <  i \leq 2s,
\end{cases}
\end{align}
for $i \in [d]$. 
This result clearly shows that the convergence of the learning process is guaranteed even if we employ low-cost low-resolution ADCs at the receiver. 

\section{DSGD with Low-Resolution DACs and ADCs}
We now consider a system where the workers and the PS employ low-resolution DACs and ADCs, respectively. 
Each worker uses a finite resolution DAC to quantize the OFDM words, and transmits them through a multipath fading channel. 
The PS receives the signal from multiple antennas where finite resolution ADCs are employed at each receive chain. 
The aim is to obtain an estimate of the gradients using the received signals, which are distorted by ADCs and DACs as well as the multipath fading channel impairments. 
%
%
We analyze the impact of employing finite resolution ADCs and DACs jointly on the convergence of the learning algorithm.
We accomplish this by using the Bussgang decomposition and AQNM model for the quantization operation at the workers and the PS, respectively.

Each worker calculates their local gradients and their corresponding OFDM words $\mathbf{\bar{G}}_m \in \mathbb{C}^{N+N_{cp}}$. 
As in Section \ref{fin_dacs}, each worker uses a finite resolution DAC, and quantizes the OFDM words corresponding to the local gradients. The $n$-th element of the transmitted signal by the $m$-th worker is given by
\begin{equation} \label{aqnm_eq2}
{\bar{G}}_m^Q[n] = Q({\bar{G}}_m[n]) = (1-\eta){\bar{G}}_m[n] + {q_m[n]}
\end{equation}
using the Bussgang decomposition. Here $\eta = 1/\text{SQNR}$ due to the quantization of ${\bar{G}}_m[n]$, and the variance of the distortion noise is $ \sigma_{q_m}^2 =   \eta (1-\eta) \sigma_{G_m}^2$.

The quantized signals pass through a multipath fading channel whose impulse response is given in \eqref{ch_impulse}. 
After removing the CP, the received signal at the input of the finite resolution ADC of the $k$-th antenna of the PS is 
\begin{equation}
U_k[n] = \sum_{m=1}^M \sum_{l=1}^L h_{mkl} \left( (1-\eta){{G}}_m[n-\tau_{mkl}] + {q_m[n-\tau_{mkl}]}\right)+ z_k[n].
\end{equation}
The mean of $U_k[n]$ is zero, and its variance is given by
\begin{align} 
\sigma_{U_k}^2 &= \sum_{m=1}^M  \sum_{l =1}^L |h_{mkl}|^2 \left( (1-\eta)^2 + \eta(1-\eta)\right) \sigma_{G_m}^2 \nonumber \\
  &  \ \ \ \ \  \ \  \ \ \ +(1-\eta)^2 \sum_{m=1}^M  \sum_{l =1}^L \sum_{l' =1, l' \not= l }^L  h_{mkl} h_{mkl'}^* \mathbb{E} \bigg[G_m[n-\tau_{mkl}] G_m[n-\tau_{mkl'}] \bigg] +\sigma_z^2,
\end{align}
which only depends on the receive antenna index $k$.

The PS employs finite resolution ADCs at each receive antenna.
Similar to quantization with DAC, the quantization operation of the ADC can be modeled as a linear operation using an AQNM model where the correlation of distortion noise across the antennas is ignored.  
The corresponding quantized signal at the $k$-th antenna is written as
\begin{equation} \label{Rk_both}
R_k[n]  = (1-\eta_k) \bigg(\sum_{m=1}^M \sum_{l=1}^L h_{mkl}  (1-\eta){{G}}_m[n-\tau_{mkl}] + \sum_{m=1}^M \sum_{l=1}^L h_{mkl}{q_m[n-\tau_{mkl}]} + z_k[n]\bigg) 
+ v_q[n],
\end{equation}
where $\eta_k$ is the distortion factor due to quantization of the received signal at the $k$-th antenna ($\mathbf{{U}}_k$), and calculated through the SQNR of the corresponding quantization operation as $\eta_k = 1/\text{SQNR}$. 
$v_q[n]$ is a non-Gaussian distortion noise whose variance is  $\sigma_{v_q}^2 = \eta_k (1-\eta_k)\sigma_{U_k}^2$.

The total effective non-Gaussian noise due to the channel and quantization with ADC at the PS is
\begin{equation}
p_k[n] =(1-\eta_k) z_k[n]
+ v_q[n],
\end{equation}
with variance $\sigma_{p_k}^2 = (1-\eta_k)^2 \sigma_z^2 + \sigma_{v_q }^2$, and
the output of the complex ADC can be rewritten as
\begin{align} \label{Yqk_fr_all}
R_k[n] = (1&-{\eta}_k)(1-\eta) \sum_{m=1}^M \sum_{l=1}^L h_{mkl} G_m[n-\tau_{mkl}] \nonumber \\
&+ (1-\eta_k) \sum_{m=1}^M \sum_{l=1}^L h_{mkl}{q_m[n-\tau_{mkl}]} + p_k[n].
\end{align}

For demodulation, we take the DFT of (\ref{Yqk_fr_all}) which results in
\begin{equation} \label{fft_both}
r_k[i] = (1-\eta_k)(1-\eta) \sum_{m=1}^M H_{mk}[i] g_m[i] +  (1-\eta_k)\sum_{m=1}^M H_{mk}[i] Q_m[i] + P_k[i],
\end{equation}
where  $H_{mk}[i]$'s are defined by \eqref{Hmk}, and $Q_m[i]$ is the DFT of the quantization distortion noise. 

Taking DFT of the effective noise, $P_k[i]$ is evaluated as
$
P_k[i] = \sum_{n = 0}^{N-1} p_k[n]e^{-j2\pi i n/N}. 
$
With a similar approach to the one used in Section \ref{fin_adcs}, $P_k[i]$ converges absolutely to a Gaussian random variable by an application of CLT \cite{billingsley}.

Since the CSI is only available at the PS as in \cite{amiri}, the received signals can be combined to align the gradient vectors as
\begin{equation} \label{yQ_fr_both}
y[i] = \frac{1}{K}\sum_{k=1}^K \frac{1}{(1-\eta)(1-\eta_k)}\bigg(\sum_{m=1}^M (H_{mk}[i])^*\bigg) r_k [i].
\end{equation}
This quantity can be written as the sum of
five different terms as in Section \ref{fin_dacs}:
\begingroup
\allowdisplaybreaks
\begin{subequations} \label{y_all_both}
	\begin{align} 
	y [i] =& \underbrace{\frac{1}{K} \sum_{k=1}^K \sum_{m=1}^M |H_{mk}[i]|^2g_m[i]}_{\text{signal term}} \\  
	&+ \underbrace{\frac{1}{K} \sum_{k=1}^K \sum_{m=1}^M \sum_{\substack{m'=1 \\ m' \neq m}}^M (H_{mk}[i])^* H_{m'k}[i] g_{m'}[i]}_{\text{interference term}} \\  
	&+ \underbrace{\frac{1}{(1-\eta) K} \sum_{k=1}^K   \sum_{m=1}^M \sum_{\substack{m'=1 \\ m' \neq m}}^M (H_{mk}[i])^* H_{m'k}[i] Q_{m'}[i]}_{\text{distortion noise term}} \\  
	&+ \underbrace{\frac{1}{(1-\eta) K} \sum_{k=1}^K \sum_{m=1}^M  |H_{mk}[i]|^2 Q_{m'}[i]}_{\text{second type of distortion noise term }} \\  
	&+ \underbrace{ \frac{1}{(1-\eta) K} \sum_{k=1}^K \frac{1}{(1-\eta_k)} \bigg(\sum_{m=1}^M (H_{mk}[i])^*\bigg) P_k[i]}_{\text{noise term}},
	\end{align}
\end{subequations}
\endgroup
which are the same as the terms given in \eqref{y_all11} except for the last noise term. As in Section \ref{fin_adcs}, the noise term, $P_k[i]$, includes both the channel noise and the quantization noise due to ADCs, and it is with zero mean and finite variance.
The analyses of the interference term (\ref{y_all_both}b), distortion noise term (\ref{y_all_both}c), and the second type of distortion noise term (\ref{y_all_both}d) are the same as those of (\ref{y_all11}b), (\ref{y_all11}c), and (\ref{y_all11}d), respectively. 
Hence, similar arguments on the convergence of the learning algorithm with finite resolution DAC are also valid for the combined effects of DACs and ADCs.  
In other words, using a sufficiently large number of antennas at the PS, the gradients can be recovered via \eqref{recover}. 
The main conclusion is that we can design a federated learning system with a large number of workers and receive antennas, and still have extremely low hardware cost and energy consumption. 
This is remarkable since it shows the practicality of the federated learning over realistic wireless channels with very low-cost hardware.

	\section{Numerical Examples} \label{sim}

	We now evaluate the performance of blind federated learning with realistic channel effects and hardware limitations via simulations. 
	Our main objective is to verify that the theoretical expectations on the low-cost federated learning systems over wireless channels are also valid via simulations. 
	We use the MNIST dataset \cite{mnist} with $60000$ training and $10000$ test samples to train a single layer neural network using the Adam optimizer \cite{kingma2014adam}. 
	At the beginning of the training process, each worker caches $B = 1000$ training samples randomly. The number of parameters is $d = 7850$.

	Our system consists of $M = 20$ workers connected to a PS through a multipath fading channel with $L = 3$ taps and $\sigma^2_{h,l} = 1/L$, hence we have a normalized uniform multipath delay profile where each tap experiences Rayleigh fading. 
	We consider an OFDM setup with $f_c = 3$ $\si{GHz}$ carrier frequency, and the number of subcarriers is $N_{\text{cp}} = 1024$ where the subcarrier spacing is $\Delta f = 80$ $\si{kHz}$. 
	We take the sampling period as $T_s = T_w/N$ where $T_w = \frac{1}{\Delta f} = 12.5 $ $\si{\micro\second}$ is the OFDM word duration without the CP. 
	As given in \cite{lee1998mobile}, the maximum delay spread of a typical urban area is $3.5 $  $\si{\micro\second}$. Consider a wireless network in an urban area where the delay spread is $3.05$  $\si{\micro\second}$ which is approximately $1000T_s$. 
	We assume that first tap has no delay and coherence time corresponds to $1000T_s$. 
	Also, time delays are uniformly spaced, i.e., $\tau_{mk1} = 0$, $\tau_{mk2} = 500T_s$, $\tau_{mk3} = 1000T_s$ for $\forall m,k$\footnotemark. \footnotetext{We select this multipath delay profile for the ease of illustration and reproduction. More realistic multipath delay profiles, e.g., uniformly distributed time delays, can be selected, but doing so will not change our main conclusions.}The cyclic prefix length is set to $N_{\text{cp}} = 1024$, which is enough to remove the ISI effects caused by the multipath. 
	The average transmit power of the OFDM word transmitted by the $m$-th worker is calculated as $P_T = \frac{1}{T}\sum_{t=1}^T \left|\left|\mathbf{\bar{G}}_m^t\right|\right|_2^2$, which gives $P_T =1.3267\times 10^{-4}$ for this setup, where $T$ is the total iteration count. 
	In our theoretical analysis, we model the OFDM words with the autocorrelation matrix $\mathbf{C}_{\mathbf{\bar{G}}_m \mathbf{\bar{G}}_m}$ with equal nonzero diagonal elements denoted by $\sigma_G^2$, and zero off-diagonal elements. 
	In our simulations, we do not make any assumption on the statistics of the gradients; we simply use the information which we obtain from our simulations to model the statistics of the gradients.

	 \begin{figure} 
	 	\centering
	 	\begin{subfigure}{0.8\textwidth} 
	 		\includegraphics[width=\textwidth]{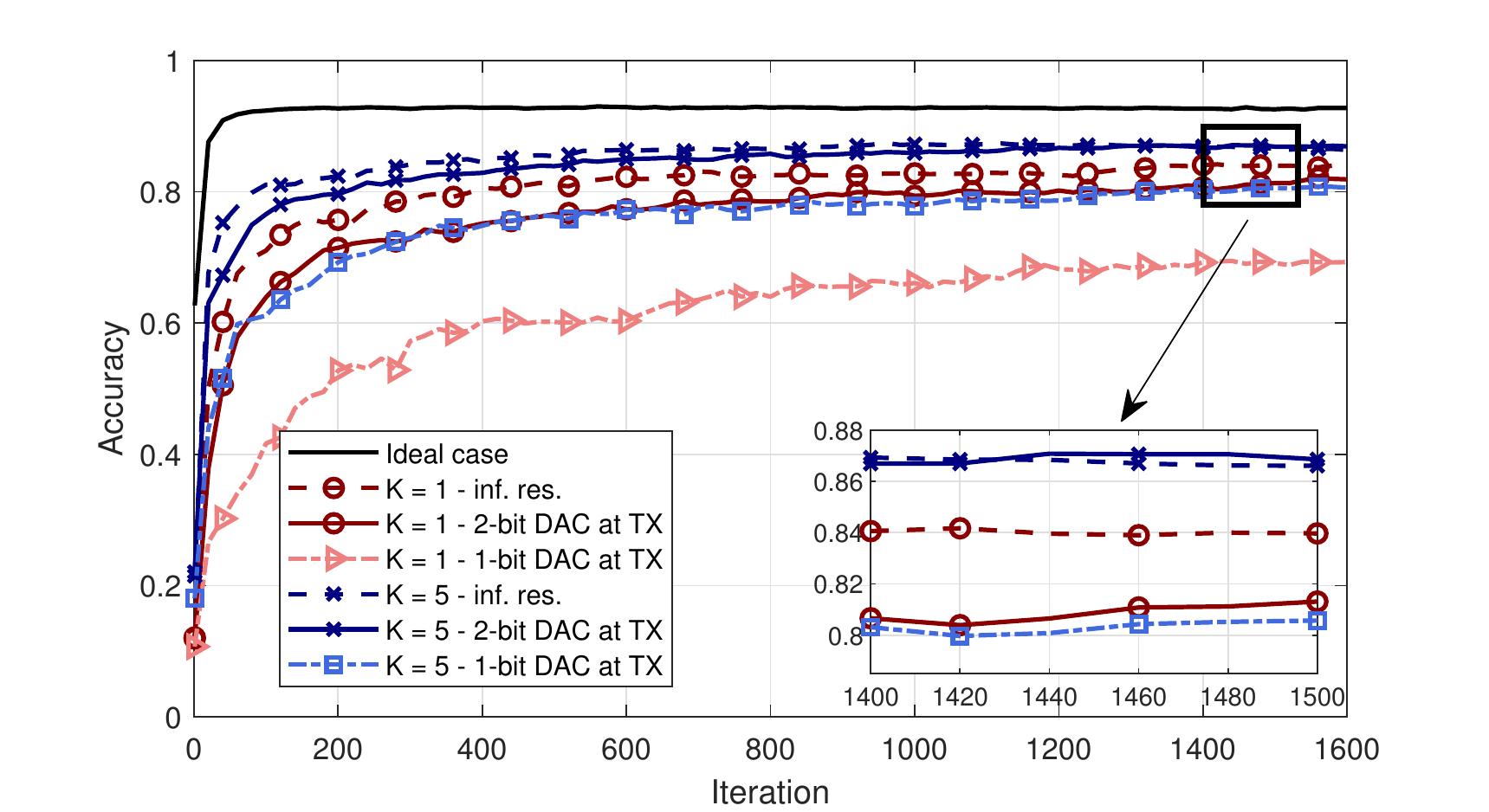}
	 		\caption{Number of receive antennas $K = 1, 5$.} 
	 	\end{subfigure}
	 	\vspace{1em} 
	 	\begin{subfigure}{0.8\textwidth} 
	 		\includegraphics[width=\textwidth]{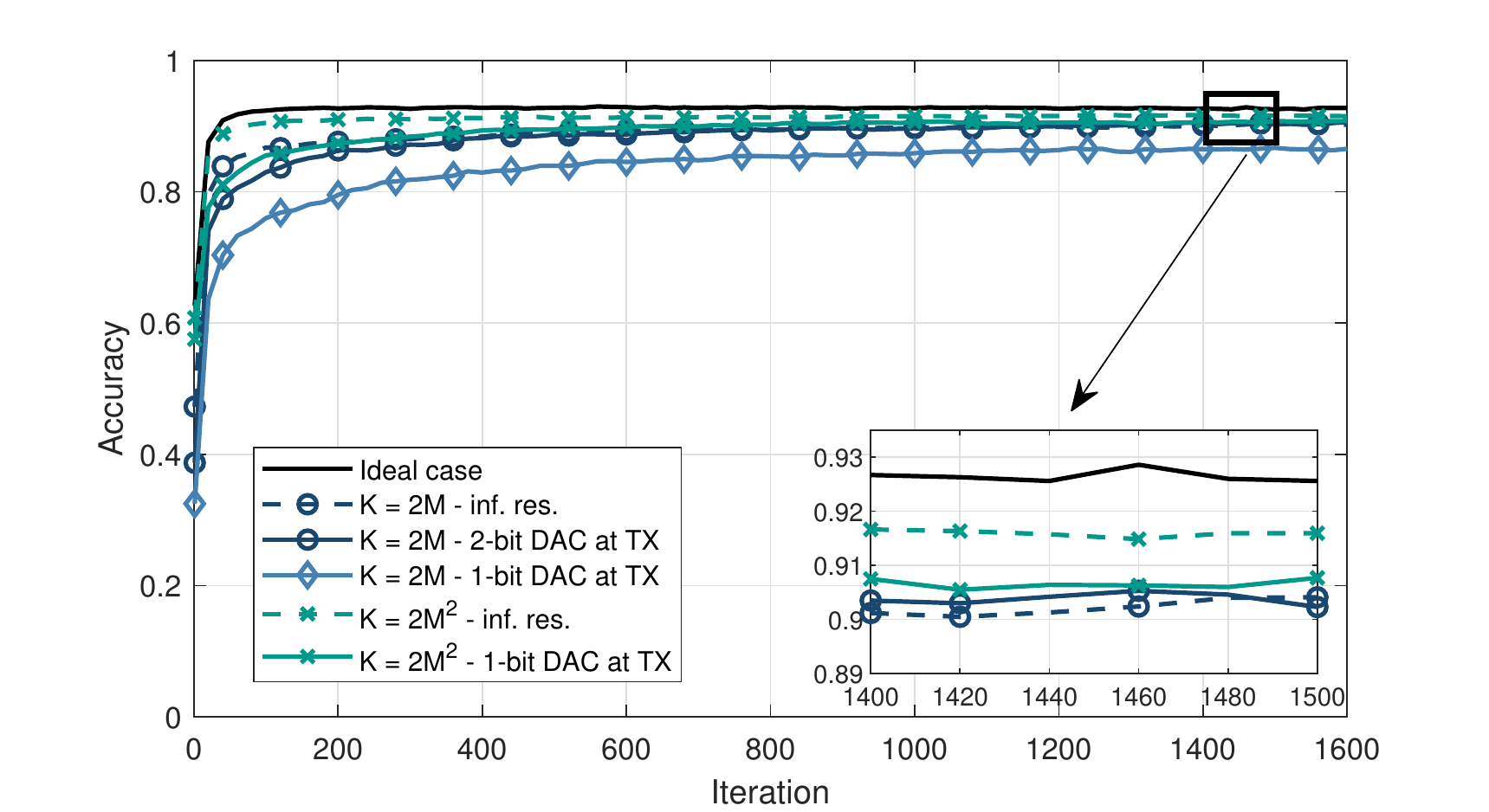}
	 		\caption{Number of receive antennas $K = 2M, 2M^2$.} 
	 	\end{subfigure}
	 	\caption{Test accuracy of the system with low-resolution DAC and channel noise variance $\sigma_z^2 = 8\times 10^{-4}$.}\label{dac12} 
	 \end{figure}

 	 \begin{figure} 
 	\centering
 	\begin{subfigure}{0.8\textwidth} 
 		\includegraphics[width=\textwidth]{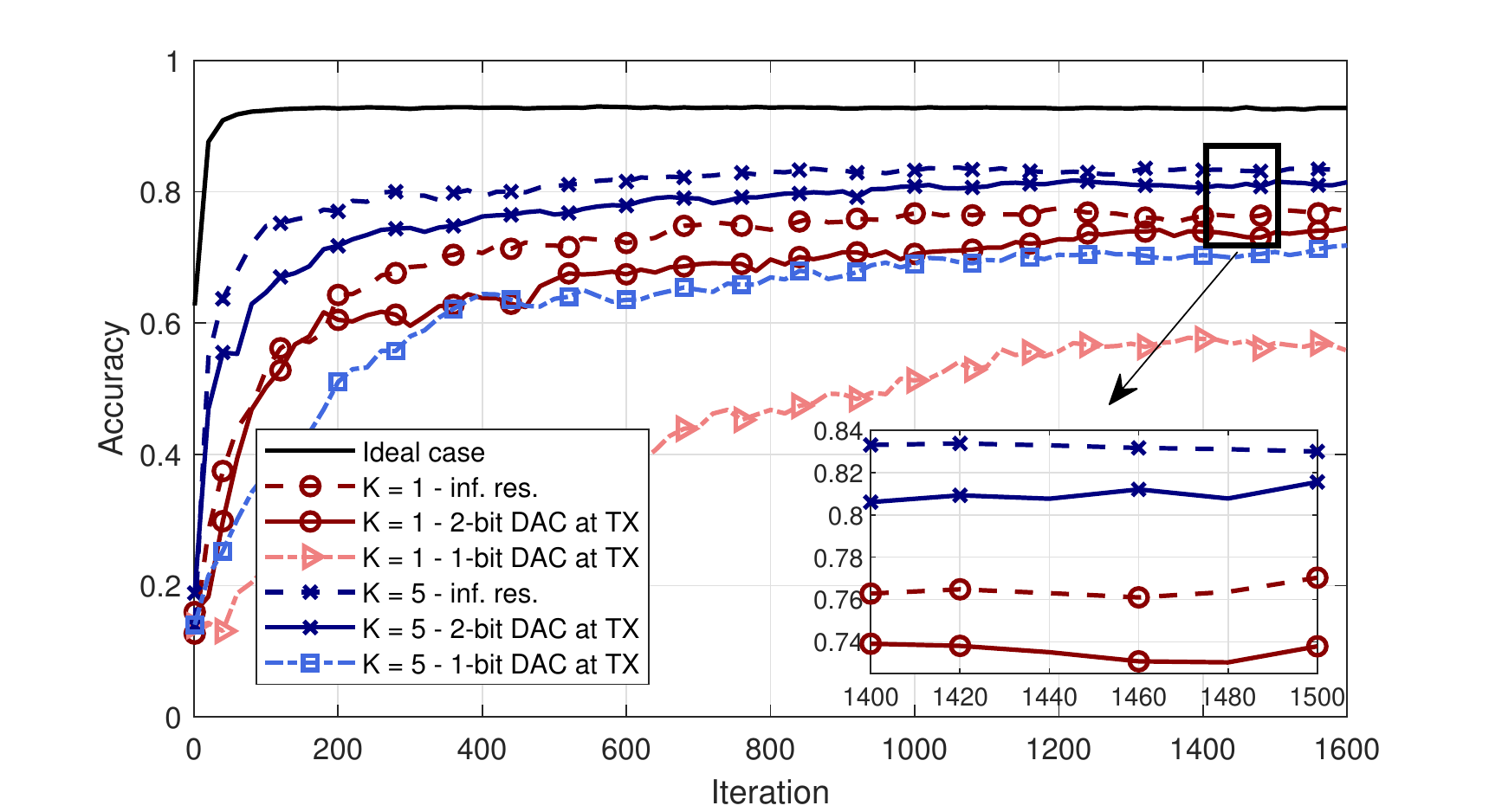}
 		\caption{Number of receive antennas $K = 1, 5$.} 
 	\end{subfigure}
 	\vspace{1em} 
 	\begin{subfigure}{0.8\textwidth} 
 		\includegraphics[width=\textwidth]{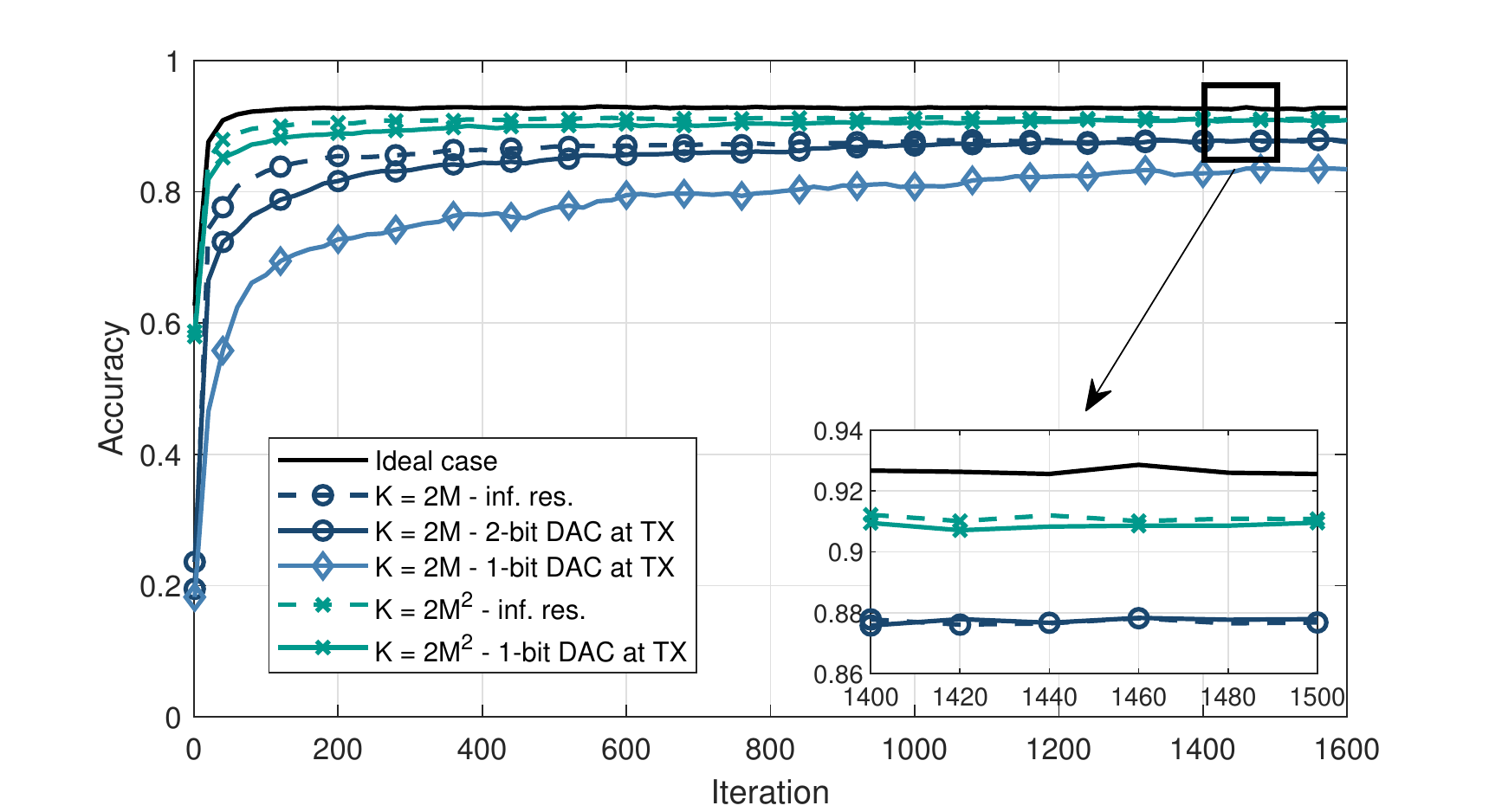}
 		\caption{Number of receive antennas $K = 2M, 2M^2$.} 
 	\end{subfigure}
 	\caption{Test accuracy of the system with low-resolution DAC and channel noise variance $\sigma_z^2 = 4\times 10^{-3}$.} \label{dac34}
 \end{figure}

	 	 \begin{figure}
		\centering
		\begin{subfigure}{0.8\textwidth} 
			\includegraphics[width=\textwidth]{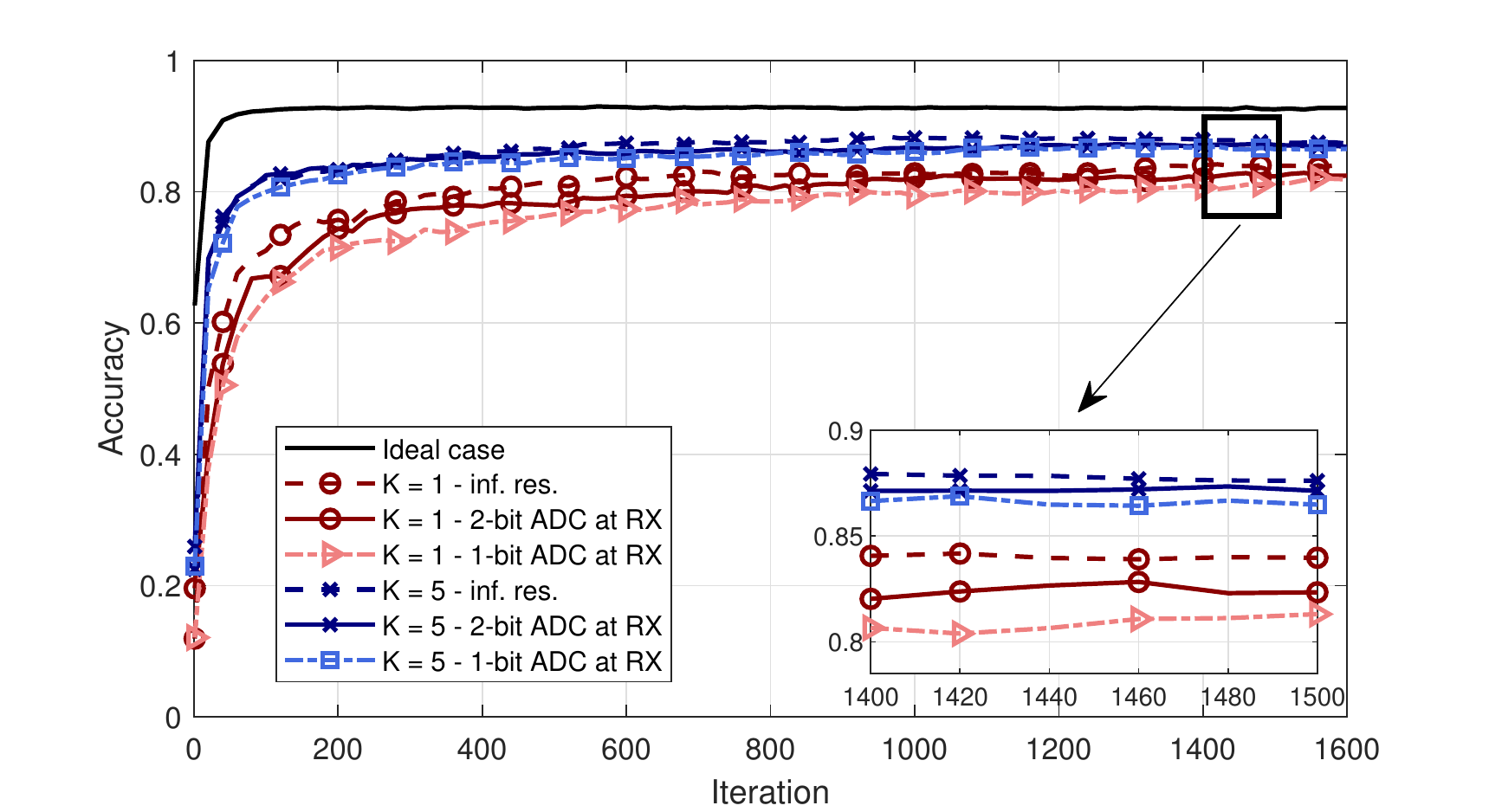}
			\caption{Number of receive antennas $K = 1, 5$.} 
		\end{subfigure}
		\vspace{1em} 
		\begin{subfigure}{0.8\textwidth} 
			\includegraphics[width=\textwidth]{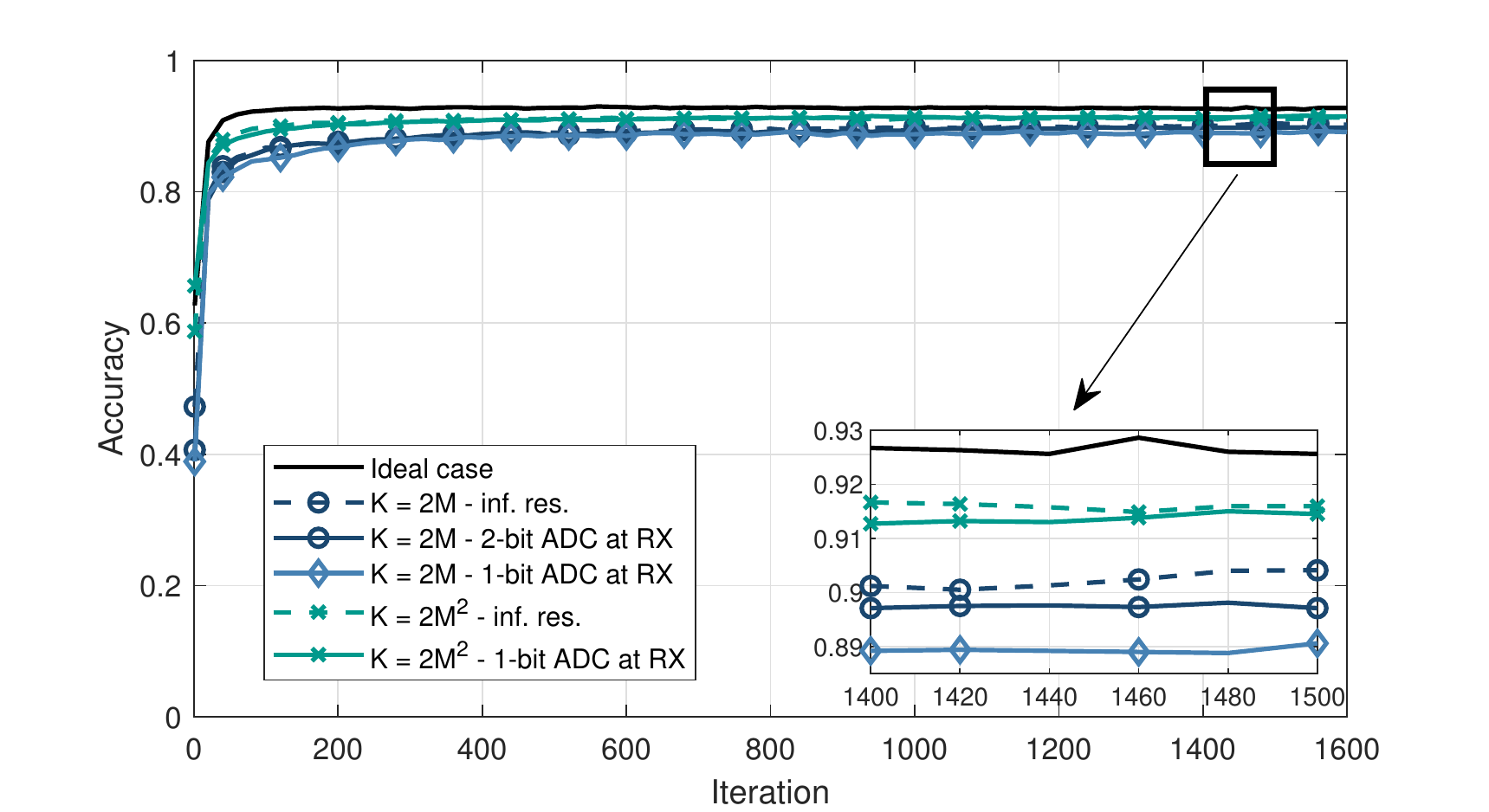}
			\caption{Number of receive antennas $K = 2M, 2M^2$.} 
		\end{subfigure}
		\caption{Test accuracy of the system with low-resolution ADC and channel noise variance $\sigma_z^2 = 8\times 10^{-4}$.}  \label{adc12} 
	\end{figure}

	 	 \begin{figure} 
	\centering
	\begin{subfigure}{0.8\textwidth} 
		\includegraphics[width=\textwidth]{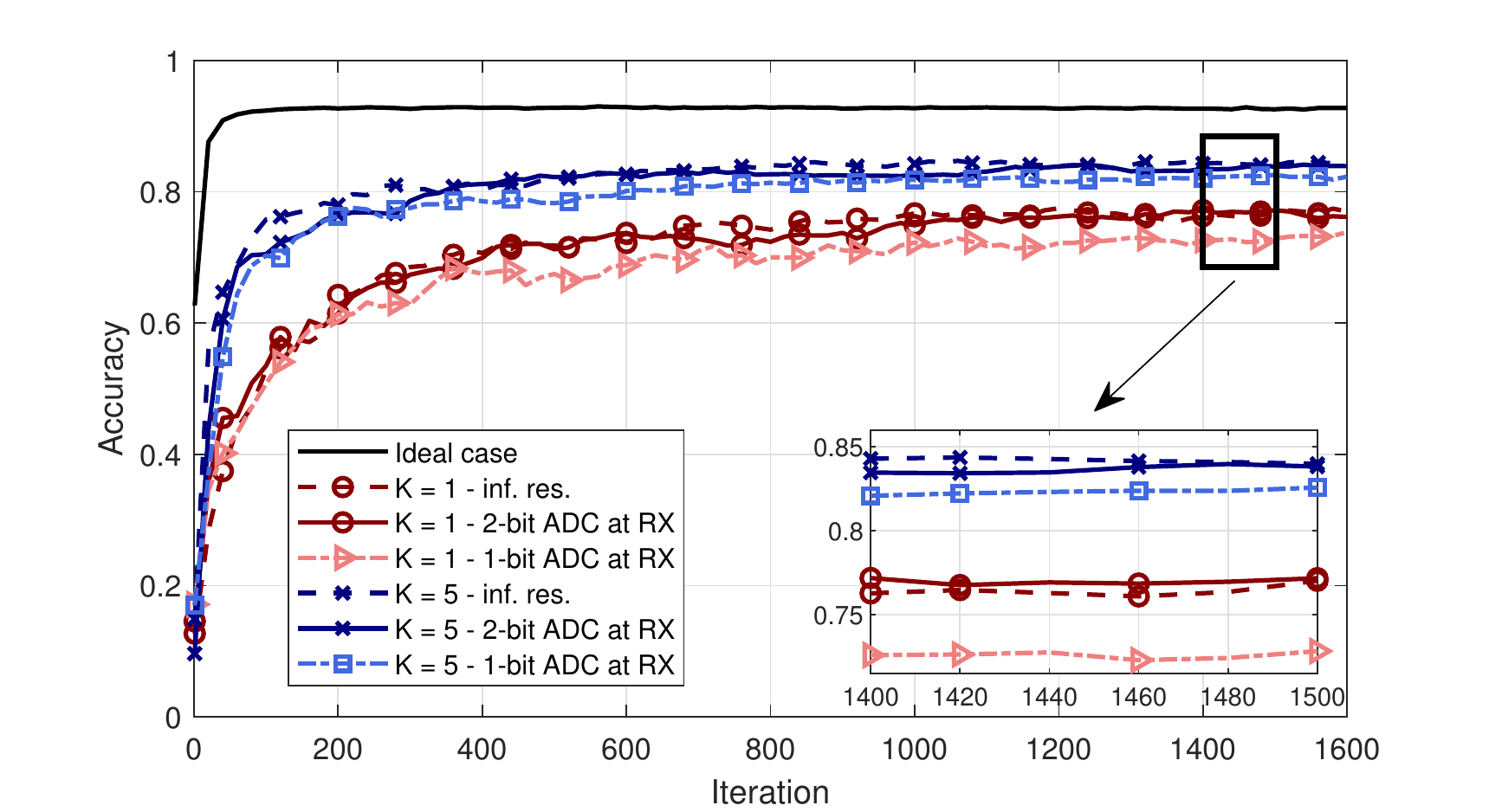}
		\caption{Number of receive antennas $K = 1, 5$.} 
	\end{subfigure}
	\vspace{1em} 
	\begin{subfigure}{0.8\textwidth} 
		\includegraphics[width=\textwidth]{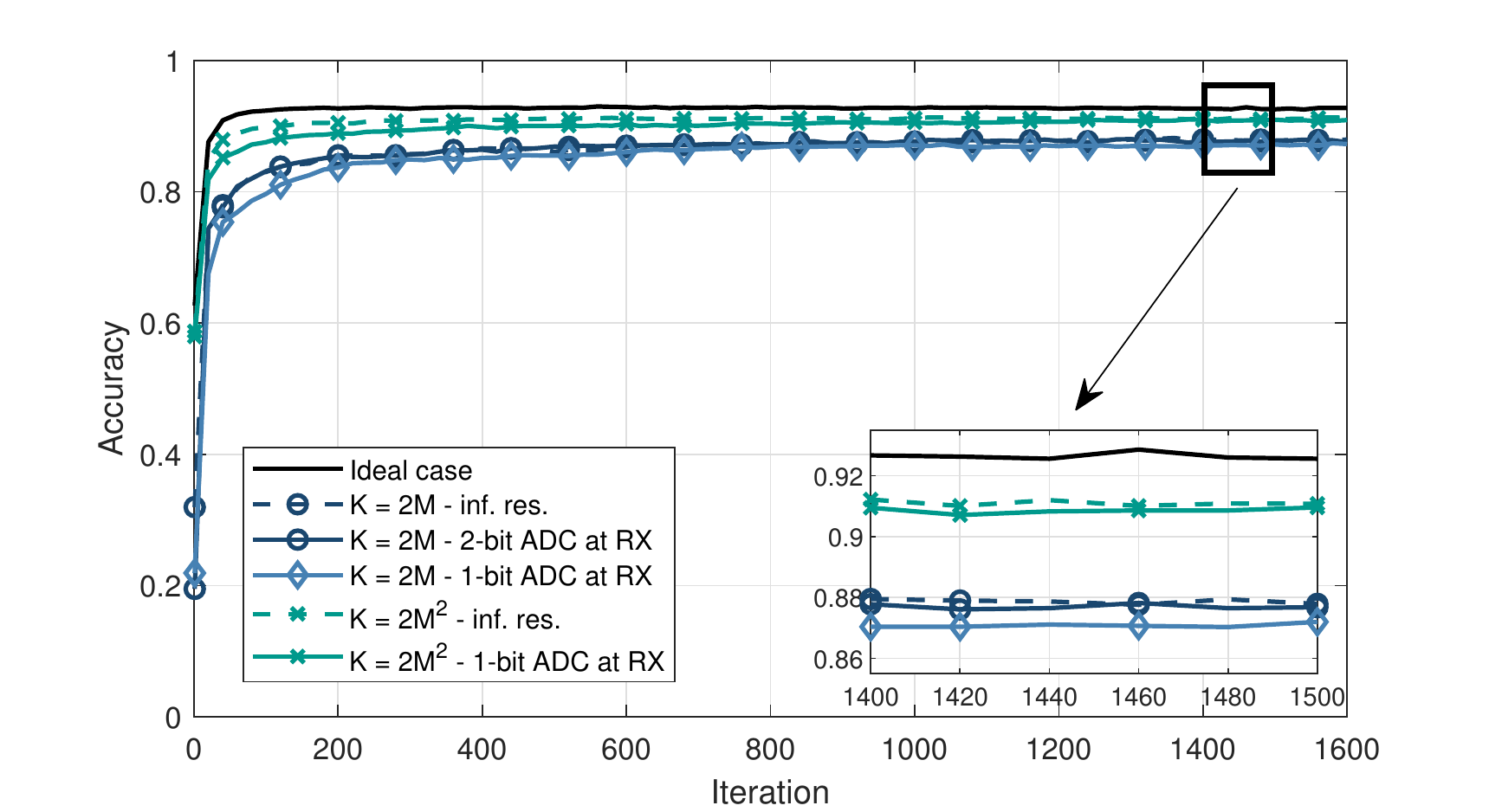}
		\caption{Number of receive antennas $K = 2M, 2M^2$.} 
	\end{subfigure}
	\caption{Test accuracy of the system with low-resolution ADC and channel noise variance $\sigma_z^2 = 4\times 10^{-3}$.} \label{adc34} 
\end{figure}

	 	 \begin{figure}
	\centering
	\begin{subfigure}{0.8\textwidth} 
		\includegraphics[width=\textwidth]{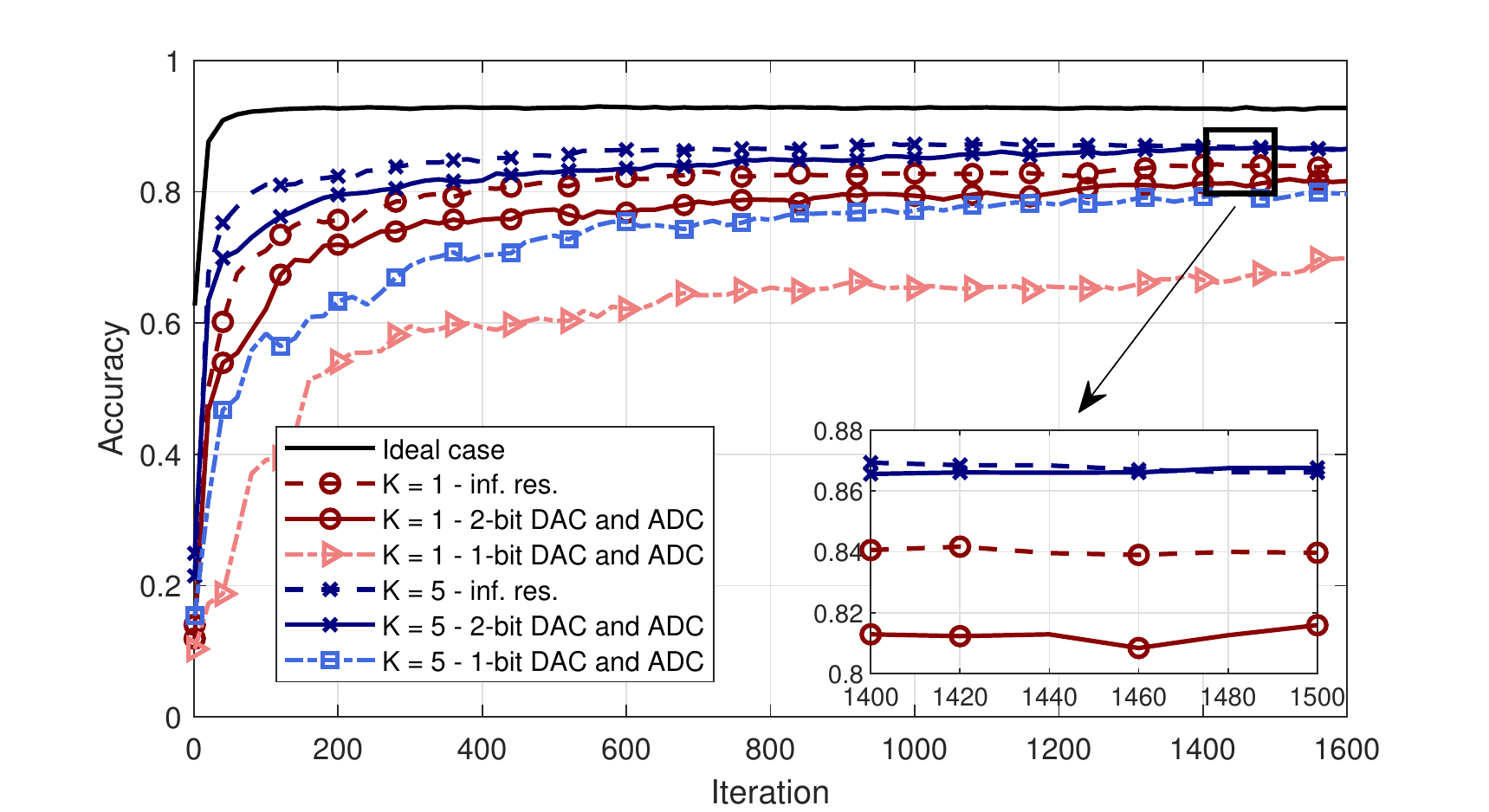}
		\caption{Number of receive antennas $K = 1, 5$.} 
	\end{subfigure}
	\vspace{1em} 
	\begin{subfigure}{0.8\textwidth} 
		\includegraphics[width=\textwidth]{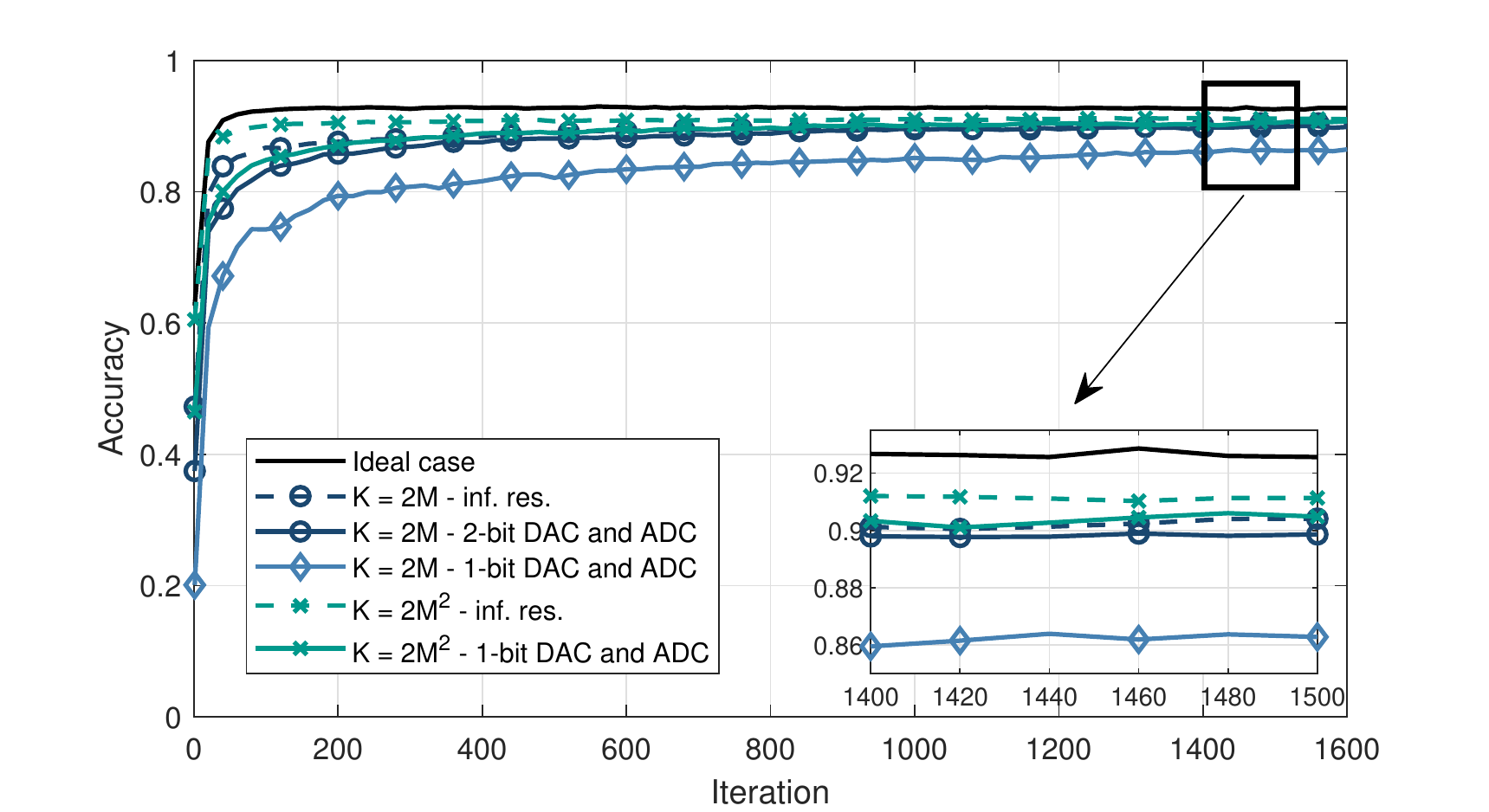}
		\caption{Number of receive antennas $K = 2M, 2M^2$.} 
	\end{subfigure}
	\caption{Test accuracy of the system with low-resolution DAC and ADC and channel noise variance $\sigma_z^2 = 8\times 10^{-4}$.}  \label{adc_dac12} 
\end{figure}

	 	 \begin{figure} 
	\centering
	\begin{subfigure}{0.8\textwidth} 
		\includegraphics[width=\textwidth]{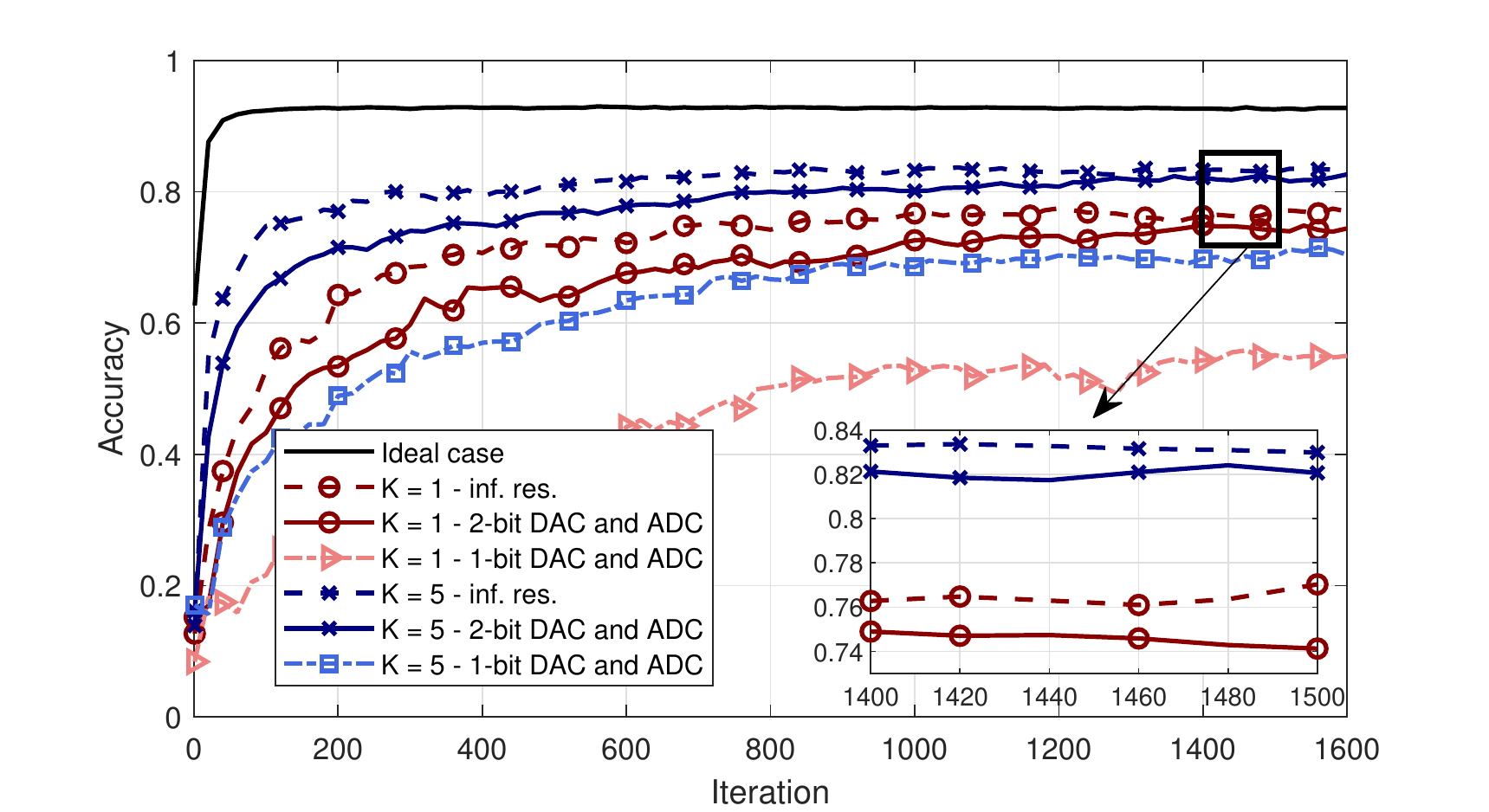}
		\caption{Number of receive antennas $K = 1, 5$.} 
	\end{subfigure}
	\vspace{1em} 
	\begin{subfigure}{0.8\textwidth} 
		\includegraphics[width=\textwidth]{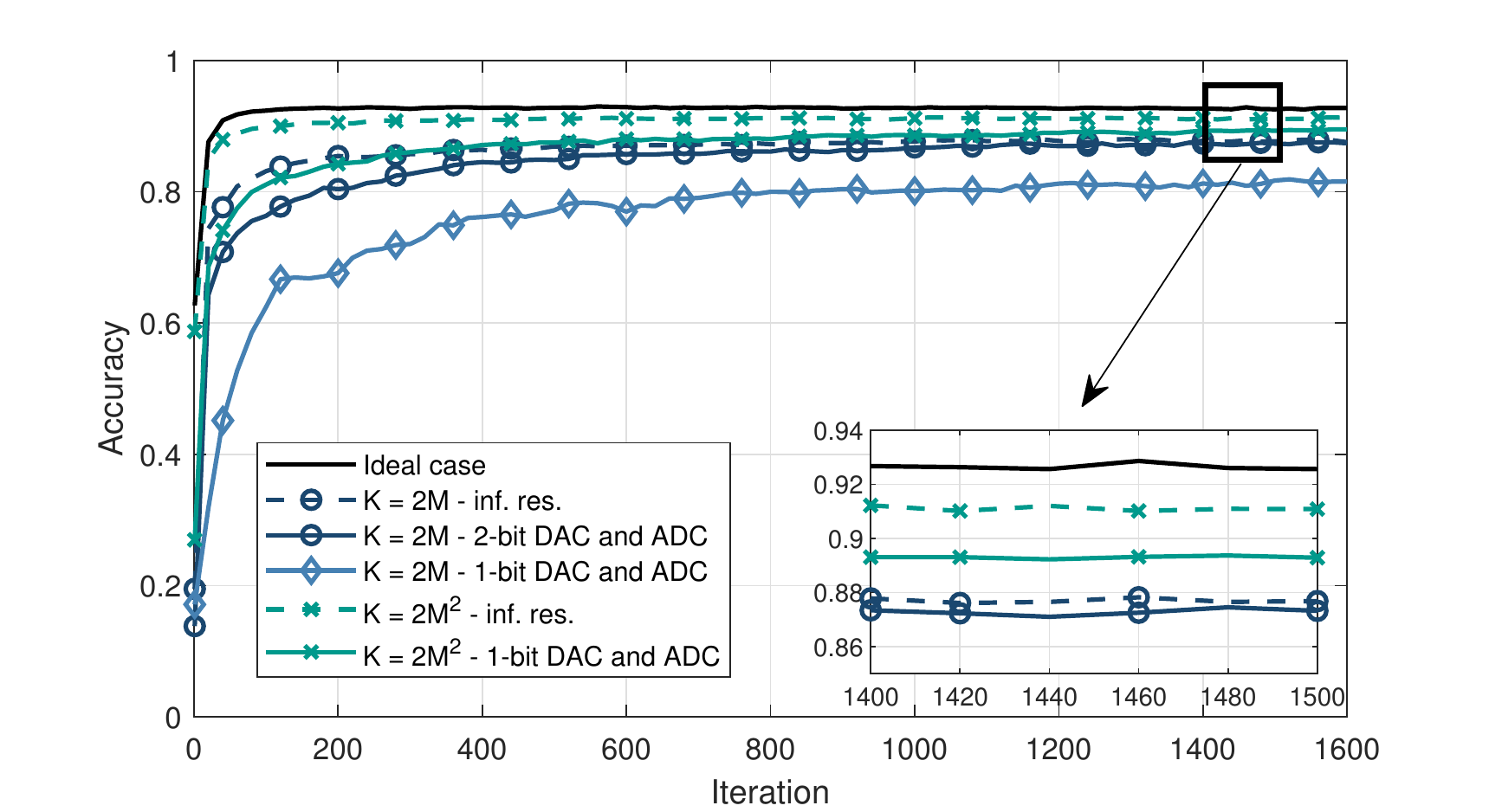}
		\caption{Number of receive antennas $K = 2M, 2M^2$.} 
	\end{subfigure}
	\caption{Test accuracy of the system with low-resolution DAC and ADC and channel noise variance $\sigma_z^2 = 4\times 10^{-3}$.} \label{adc_dac34} 
\end{figure}

\begin{figure}[t]
	\centering
	\includegraphics[width=0.8\textwidth]{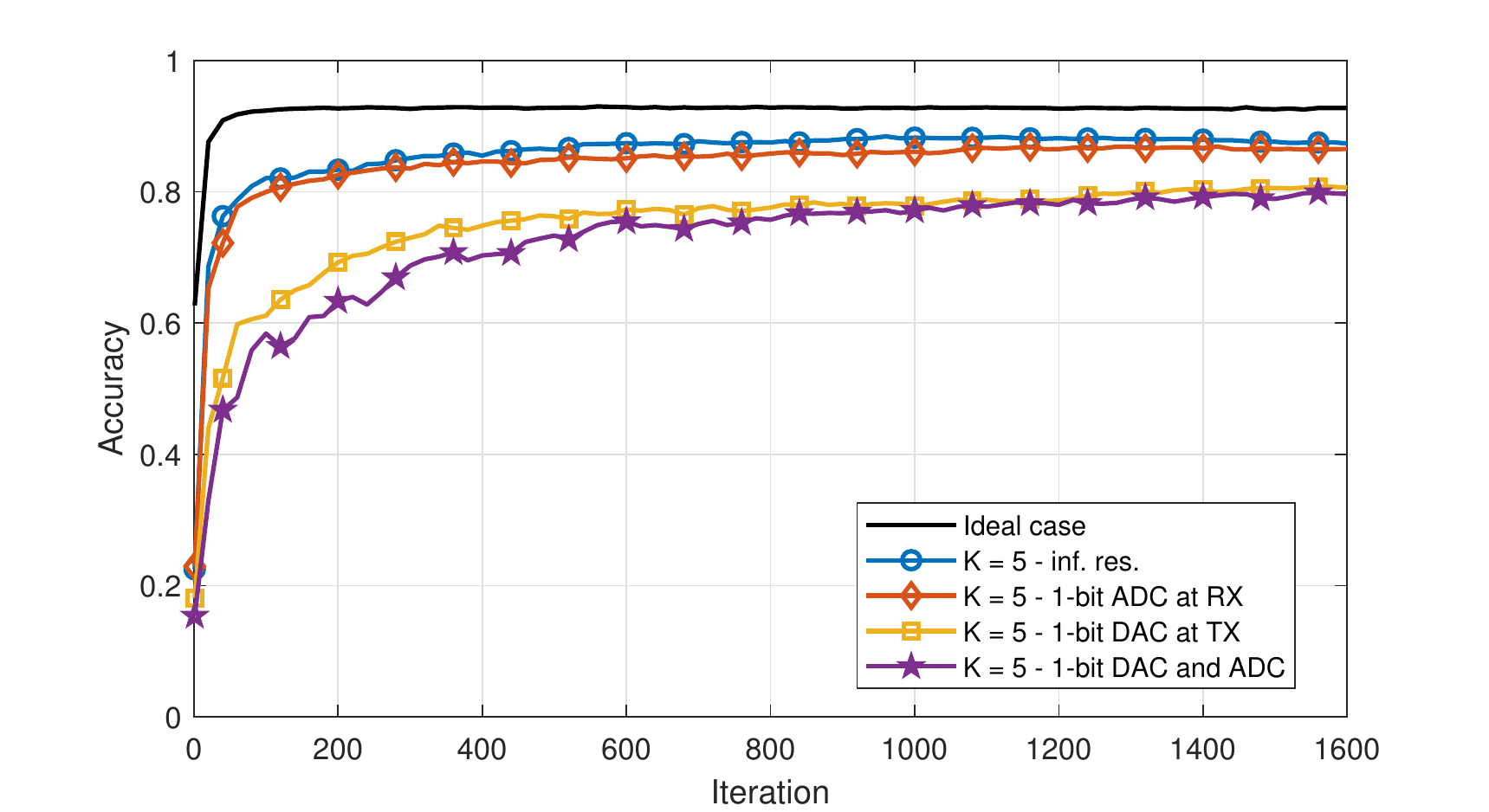}
	\caption{Test accuracy of the system with seperate one-bit DACs at the workers, one-bit ADCs at the PS antennas, and joint DACs and ADCs where the channel noise variance is $\sigma_z^2 = 8\times 10^{-4}$, and $K = 5$. }  \label{comparison}
\end{figure}

	In Figs. \ref{dac12}a and \ref{dac12}b, the test accuracy for a system where each worker is equipped with low-resolution DAC and  different number of antennas $K \in \{1,\ 5,\ M,\ 2M^2\}$ at the receiver side is illustrated for a system with $\sigma_z^2 = 8\times 10^{-4}$. 
	As the number of receive antennas increases, the test accuracy approaches that of the infinite resolution case since the variance of the distortion noise and interference decreases.
	At iteration $T = 1600$, the accuracy loss with one-bit DAC compared to infinite resolution case is $17.62\%$, $6.62\%$, $ 4.07\%$, and $0.37\%$ for $K = 1$, $K = 5$, $K = 2M$, and $K = 2M^2$, respectively. 
	Furthermore, the low complexity system achieves almost the same accuracy with infinite resolution case when two-bit DACs are employed (except for $K=1$ which has an accuracy loss of $2.64\%$).
	In Fig. \ref{dac34}a and \ref{dac34}b, we increase the channel noise variance to $\sigma_z^2 = 4\times 10^{-3}$, i.e., there is $14$ dB SNR reduction. 
	As expected, the performance of the learning algorithm deteriorates, since the effect of noise term is increased. 
	However, as shown in Figs. \ref{dac34}a and \ref{dac34}b, the convergence is still achieved, and the accuracy loss of one-bit DAC case compared to infinite resolution case is $27.54\%$, $13.95\%$, $ 4.71\%$, and $0.8\%$ for $K = 1$, $K = 5$, $K = 2M$, and $K = 2M^2$, respectively. 
	With two-bit DACs, the accuracy loss decreases to $3.26\%$ and $2.40\%$ for $K = 1$ and $K = 5$, respectively, while it gives almost same performance when the number of PS antennas is $K = 2M$ and $K = 2M^2$. 
	These results clearly illustrate that when moderate number of receive antennas are employed, low-resolution, even two-bit, DACs can achieve a learning performance comparable with the infinite resolution case.

	In Figs. \ref{adc12}a and \ref{adc12}b, the test accuracy for different number of antennas $K \in \{1,\ 5,\ M,\ 2M^2\}$ each equipped with a low-resolution ADC is illustrated for a system with $\sigma_z^2 = 8\times 10^{-4}$, and compared with the error-free shared link case. 
	As expected, using higher number of receive antennas results in an improved learning accuracy. 
	Indeed the results are very close to those of the infinite resolution case, especially with two-bit ADCs, while there is a minor drop on accuracy with one-bit ADCs.
	For instance, after the $1600$-th iteration, using one-bit ADCs causes only $2.64\%$, $0.95\%$, and $0.13\%$, accuracy loss compared to infinite resolution case for $K = 1$, $K = 5$, and $K = 2M$ respectively. 
	Further, it achieves the performance of the infinite resolution with $K = 2M^2$ PS antennas. 
	These results are due to the fact that increasing the number of antennas reduces the interference dramatically which makes the combined signal a very good estimate of the gradient vector, even with low-resolution ADCs.

	Without changing any other parameters of the setup described above, we increase the noise variance to $\sigma_z^2 = 4\times 10^{-3}$ in Figs. \ref{adc34}a and \ref{adc34}b. 
	As in the previous case, for the two-bit ADC case, the performance of the proposed scheme is very close to the error-free case for a large number of receive antennas. 
	When the number of antennas is decreased, with the detrimental effects of the channel noise and interference caused by the multipath fading channel, the accuracy decreases. 
	However, even for this high level of channel noise, using one-bit ADCs causes only $4.09\%$, $2.55\%$, $0.37\%$, and $0.32\%$ accuracy loss compared to infinite resolution case for $K = 1$, $K = 5$, $K = 2M$, and $K = 2M^2$, respectively, after the $1600$-th iteration.

	In Figs. \ref{adc_dac12}a and \ref{adc_dac12}b, we consider a  system which employs both low-resolution DACs at the workers and one-bit ADCs at the PS antennas with channel noise variance $\sigma_z^2 = 8\times 10^{-4}$. 
	As expected, using low-resolution DAC and ADC at the same time increases the amount of interference in the gradient estimate at the PS, which decreases the learning accuracy of the distributed system. 
	However, the combined effect of the interference terms is still negligible, especially for sufficiently large number of receive antennas. 
	After the $1600$-th iteration, using one-bit DACs and ADCs simultaneously causes only $17.91\%$, $7.76\%$, $4.18\%$, and $0.39\%$ accuracy loss compared to infinite resolution case for $K = 1$, $K = 5$, $K = 2M$, and $K = 2M^2$, respectively. %
	When $K=1$, using two-bit DAC and ADC results in a $2.95\%$ accuracy loss while it is almost same as the infinite resolution case when the number of PS antennas is higher. 
	In the same system, we increase the channel noise variance to  $\sigma_z^2 = 4\times 10^{-3}$. As shown in Figs. \ref{adc_dac34}a and \ref{adc_dac34}b, increasing noise causes $28.56\%$, $15.66\%$, $6.87\%$, and $1.91\%$ accuracy loss compared to infinite resolution case for $K = 1$, $K = 5$, $K = 2M$, and $K = 2M^2$, respectively after the $1600$-th iteration with one-bit DAC and ADC. 
	With two-bit DAC and ADC, the accuracy loss decreases to $3.35\%$ and $2.57\%$ for $K = 1$ and $K = 5$, respectively.

	Finally, in Fig. \ref{comparison}, we compare the effect of one-bit quantization on the transmitter and receiver side, both separately and jointly, with a fixed number of receive antennas $K=5$.
	As expected, the test accuracy of the system with one-bit DAC workers and infinite resolution ADCs at the PS is lower than that for the case of infinite resolution DACs at the workers and one-bit ADCs at the PS. 
	This is because, using DACs at the workers results in higher interference than using ADCs, and the performance is deteriorated. 
	However, the convergence of the learning algorithm is preserved.	
	Another important implication of our results is that even though our analysis is based on a certain assumption on the statistics of the gradients, the simulation results (which are obtained without using the Gaussian assumption on the OFDM words) are consistent with our theoretical expectations. 
	Hence, with a slight sacrifice on the accuracy rate of the learning algorithm, power and hardware efficient systems (at both transmitter and receiver sides) can be designed and implemented for distributed learning at the wireless edge over realistic wireless channels.

	\section{Conclusions} \label{conc}
	We have investigated blind federated learning at the wireless edge with OFDM based transmission and low-resolution, even one-bit, DACs and ADCs at the transmitter and receiver sides, respectively, for a practical and inexpensive system design, and reduced power consumption. 
	Our analytical results illustrate that with low-resolution DACs at the transmitter and ADCs at the receiver, the convergence
	of the distributed learning algorithms based on SGD is guaranteed when the number of receive antennas is increased as in the ideal case of infinite
	resolution DACs and ADCs. 
	Moreover, the convergence is still attained with the joint use of DACs and ADCs which reduces the implementation costs further. 
	The results are also valid for the extreme case of one-bit DACs and ADCs. 
	Through extensive numerical examples, it is also illustrated that using a moderate number of antennas with low-resolution DACs and ADCs, e.g.,
	using 5 antennas at the PS, can closely approach to the performance of the infinite resolution case. 
	It is also observed that, in
	case of low channel noise, the learning performance is decreased only slightly even for the extreme case of one-bit ADCs and DACs.

	\bibliographystyle{IEEEtran}
	\bibliography{bibs_edge}

\begin{thebibliography}{10}
\providecommand{\url}[1]{#1}
\csname url@samestyle\endcsname
\providecommand{\newblock}{\relax}
\providecommand{\bibinfo}[2]{#2}
\providecommand{\BIBentrySTDinterwordspacing}{\spaceskip=0pt\relax}
\providecommand{\BIBentryALTinterwordstretchfactor}{4}
\providecommand{\BIBentryALTinterwordspacing}{\spaceskip=\fontdimen2\font plus
\BIBentryALTinterwordstretchfactor\fontdimen3\font minus
  \fontdimen4\font\relax}
\providecommand{\BIBforeignlanguage}[2]{{%
\expandafter\ifx\csname l@#1\endcsname\relax
\typeout{** WARNING: IEEEtran.bst: No hyphenation pattern has been}%
\typeout{** loaded for the language `#1'. Using the pattern for}%
\typeout{** the default language instead.}%
\else
\language=\csname l@#1\endcsname
\fi
#2}}
\providecommand{\BIBdecl}{\relax}
\BIBdecl

\bibitem{bekkerman}
R.~Bekkerman, M.~Bilenko, and J.~Langford, \emph{Scaling up machine learning:
  Parallel and distributed approaches}.\hskip 1em plus 0.5em minus 0.4em\relax
  Cambridge University Press, 2011.

\bibitem{chilimbi}
T.~Chilimbi, Y.~Suzue, J.~Apacible, and K.~Kalyanaraman, ``Project adam:
  Building an efficient and scalable deep learning training system,'' in
  \emph{11th {USENIX} Symposium on Operating Systems Design and Implementation
  ({OSDI} 14)}, Broomfield, CO, USA, Oct. 2014, pp. 571--582.

\bibitem{amiri1}
M.~Mohammadi~Amiri and D.~G{\"u}nd{\"u}z, ``Machine learning at the wireless
  edge: Distributed stochastic gradient descent over-the-air,'' \emph{IEEE
  Transactions on Signal Processing}, vol.~68, pp. 2155--2169, Mar. 2020.

\bibitem{zhu2018low}
G.~Zhu, Y.~Wang, and K.~Huang, ``Broadband analog aggregation for low-latency
  federated edge learning,'' \emph{IEEE Transactions on Wireless
  Communications}, vol.~19, no.~1, pp. 491--506, Jan. 2019.

\bibitem{amiri2}
M.~M. Amiri and D.~G{\"u}nd{\"u}z, ``Federated learning over wireless fading
  channels,'' \emph{IEEE Transactions on Wireless Communications}, vol.~19,
  no.~5, pp. 3546--3557, May 2020.

\bibitem{yang2019energy}
Z.~Yang, M.~Chen, W.~Saad, C.~S. Hong, and M.~Shikh-Bahaei, ``Energy efficient
  federated learning over wireless communication networks,'' \emph{arXiv
  preprint arXiv:1911.02417}, 2019.

\bibitem{chen2020convergence}
M.~Chen, H.~V. Poor, W.~Saad, and S.~Cui, ``Convergence time optimization for
  federated learning over wireless networks,'' \emph{arXiv preprint
  arXiv:2001.07845}, 2020.

\bibitem{qsgd}
D.~Alistarh, D.~Grubic, J.~Li, R.~Tomioka, and M.~Vojnovic, ``{QSGD}:
  Communication-efficient {SGD} via gradient quantization and encoding,'' in
  \emph{Advances in Neural Information Processing Systems}, Long Beach, CA,
  USA, Dec. 2017, pp. 1709--1720.

\bibitem{seide}
F.~Seide, H.~Fu, J.~Droppo, G.~Li, and D.~Yu, ``1-bit stochastic gradient
  descent and its application to data-parallel distributed training of speech
  {DNN}s,'' in \emph{15th Annual Conference of the International Speech
  Communication Association}, Singapore, Sep. 2014.

\bibitem{zhao2019quantized}
S.-Y. Zhao, H.~Gao, and W.-J. Li, ``Quantized {Epoch-SGD} for
  communication-efficient distributed learning,'' \emph{arXiv preprint
  arXiv:1901.03040}, 2019.

\bibitem{amiri2020federated}
M.~M. Amiri, D.~Gunduz, S.~R. Kulkarni, and H.~V. Poor, ``Federated learning
  with quantized global model updates,'' \emph{arXiv preprint
  arXiv:2006.10672}, 2020.

\bibitem{amiri}
M.~M. {Amiri}, T.~M. {Duman}, and D.~{Gündüz}, ``Collaborative machine
  learning at the wireless edge with blind transmitters,'' in \emph{IEEE Global
  Conference on Signal and Information Processing (GlobalSIP)}, Ottowa, ON,
  Canada, Nov. 2019.

\bibitem{amiri2020blindx}
M.~M. Amiri, T.~M. Duman, D.~Gunduz, S.~R. Kulkarni, and H.~V. Poor, ``Blind
  federated edge learning,'' \emph{arXiv preprint arXiv:2010.10030}, 2020.

\bibitem{studer2016quantized}
C.~Studer and G.~Durisi, ``Quantized massive {MU-MIMO-OFDM} uplink,''
  \emph{IEEE Transactions on Commununications}, vol.~64, no.~6, pp. 2387--2399,
  Jun. 2016.

\bibitem{mollen2016uplink}
C.~Mollen, J.~Choi, E.~G. Larsson, and R.~W. Heath, ``Uplink performance of
  wideband massive {MIMO} with one-bit {ADC}s,'' \emph{IEEE Transactions on
  Wireless Communications}, vol.~16, no.~1, pp. 87--100, Jan. 2016.

\bibitem{dardari2006joint}
D.~Dardari, ``Joint clip and quantization effects characterization in {OFDM}
  receivers,'' \emph{IEEE Transactions on Circuits and Systems {I}: Regular
  Papers}, vol.~53, no.~8, pp. 1741--1748, Aug. 2006.

\bibitem{one0}
J.~Zhang, L.~Dai, X.~Li, Y.~Liu, and L.~Hanzo, ``On low-resolution {ADC}s in
  practical {5G} millimeter-wave massive {MIMO} systems,'' \emph{IEEE
  Communications Magazine}, vol.~56, no.~7, pp. 205--211, Jul. 2018.

\bibitem{one1}
S.~Jacobsson, G.~Durisi, M.~Coldrey, U.~Gustavsson, and C.~Studer, ``Throughput
  analysis of massive {MIMO} uplink with low-resolution {ADC}s,'' \emph{IEEE
  Transactions on Wireless Communications}, vol.~16, no.~6, pp. 4038--4051,
  Jun. 2017.

\bibitem{one2}
Y.~Li, C.~Tao, G.~Seco-Granados, A.~Mezghani, A.~L. Swindlehurst, and L.~Liu,
  ``Channel estimation and performance analysis of one-bit massive {MIMO}
  systems,'' \emph{IEEE Transactions on Signal Processing}, vol.~65, no.~15,
  pp. 4075--4089, Aug. 2017.

\bibitem{xu2017user}
J.~Xu, W.~Xu, F.~Shi, and H.~Zhang, ``User loading in downlink multiuser
  massive {MIMO} with 1-bit {DAC} and quantized receiver,'' in \emph{IEEE 86th
  Vehicular Technology Conference (VTC-Fall)}, Toronto, ON, Canada, Sep. 2017.

\bibitem{jacobsson2017massive}
S.~Jacobsson, G.~Durisi, M.~Coldrey, and C.~Studer, ``Massive {MU-MIMO-OFDM}
  downlink with one-bit {DAC}s and linear precoding,'' in \emph{IEEE Global
  Communications Conference}, Singapore, Dec. 2017.

\bibitem{walden1999analog}
R.~H. Walden, ``Analog-to-digital converter survey and analysis,'' \emph{IEEE
  Journal on Selelted Areas in Communications}, vol.~17, no.~4, pp. 539--550,
  Apr. 1999.

\bibitem{lee2008analog}
H.-S. Lee and C.~G. Sodini, ``Analog-to-digital converters: Digitizing the
  analog world,'' \emph{Proceedings of the IEEE}, vol.~96, no.~2, pp. 323--334,
  Feb. 2008.

\bibitem{ofdm}
S.~Wei, D.~L. Goeckel, and P.~A. Kelly, ``Convergence of the complex envelope
  of bandlimited {OFDM} signals,'' \emph{IEEE Transactions on Information
  Theory}, vol.~56, no.~10, pp. 4893--4904, Oct. 2010.

\bibitem{jacobsson2018massive}
S.~Jacobsson, U.~Gustavsson, G.~Durisi, and C.~Studer, ``Massive {MU-MIMO-OFDM}
  uplink with hardware impairments: Modeling and analysis,'' in \emph{52nd
  Asilomar Conference on Signals, Systems, and Computers}, Pacific Grove, CA,
  USA, Oct. 2018, pp. 1829--1835.

\bibitem{aghdam2019performance}
S.~R. Aghdam and T.~Eriksson, ``On the performance of distortion-aware linear
  receivers in uplink massive {MIMO} systems,'' in \emph{16th International
  Symposium on Wireless Communication Systems (ISWCS)}, Oulu, Finland, Aug.
  2019, pp. 208--212.

\bibitem{fan2015uplink}
L.~Fan, S.~Jin, C.-K. Wen, and H.~Zhang, ``Uplink achievable rate for massive
  {MIMO} systems with low-resolution {ADC},'' \emph{IEEE Communications
  Letters}, vol.~19, no.~12, pp. 2186--2189, Dec. 2015.

\bibitem{max1960quantizing}
J.~Max, ``Quantizing for minimum distortion,'' \emph{IRE Transactions on
  Information Theory}, vol.~6, no.~1, pp. 7--12, Mar. 1960.

\bibitem{bussgang}
J.~J. Bussgang, ``Crosscorrelation functions of amplitude-distorted {G}aussian
  signals,'' 1952.

\bibitem{demir2020bussgang}
{\"O}.~T. Demir and E.~Bj{\"o}rnson, ``The {Bussgang} decomposition of
  non-linear systems: Basic theory and {MIMO} extensions,'' \emph{arXiv
  preprint arXiv:2005.01597}, 2020.

\bibitem{bjornson2018hardware}
E.~Bj{\"o}rnson, L.~Sanguinetti, and J.~Hoydis, ``Hardware distortion
  correlation has negligible impact on {UL} massive {MIMO} spectral
  efficiency,'' \emph{IEEE Transactions on Communications}, vol.~67, no.~2, pp.
  1085--1098, Feb. 2018.

\bibitem{aqnm}
A.~K. Fletcher, S.~Rangan, V.~K. Goyal, and K.~Ramchandran, ``Robust predictive
  quantization: Analysis and design via convex optimization,'' \emph{IEEE
  Journal of Selected Topics in Signal Processing}, vol.~1, no.~4, pp.
  618--632, Dec. 2007.

\bibitem{orhan2015low}
O.~Orhan, E.~Erkip, and S.~Rangan, ``Low power analog-to-digital conversion in
  millimeter wave systems: Impact of resolution and bandwidth on performance,''
  in \emph{Information Theory and Applications Workshop (ITA)}, San Diego, CA,
  USA, Feb. 2015, pp. 191--198.

\bibitem{zhang2018mixed}
J.~Zhang, L.~Dai, Z.~He, B.~Ai, and O.~A. Dobre, ``Mixed-{ADC/DAC} multipair
  massive {MIMO} relaying systems: Performance analysis and power
  optimization,'' \emph{IEEE Transactions on Communications}, vol.~67, no.~1,
  pp. 140--153, Jan. 2018.

\bibitem{billingsley}
P.~Billingsley, \emph{Probability and measure}.\hskip 1em plus 0.5em minus
  0.4em\relax John Wiley \& Sons, 2008.

\bibitem{mnist}
Y.~LeCun, ``The {MNIST} database of handwritten digits,'' \emph{http://yann.
  lecun. com/exdb/mnist/}, 1998.

\bibitem{kingma2014adam}
D.~P. Kingma and J.~Ba, ``Adam: A method for stochastic optimization,''
  \emph{arXiv preprint arXiv:1412.6980}, 2014.

\bibitem{lee1998mobile}
W.~C. Lee, \emph{Mobile communications engineering: theory and
  applications}.\hskip 1em plus 0.5em minus 0.4em\relax McGraw-Hill Education,
  1998.

\end{thebibliography}
	
	\end{document}